\documentclass[reprint,amsmath,amssymb,aps,superscriptaddress,longbibliography]{revtex4-1}
\usepackage{dcolumn}
\usepackage{bm}
\usepackage{braket}
\usepackage[dvipdfmx]{graphicx}
\usepackage[dvipdfmx]{color}
\usepackage{pstricks,pst-grad,color}
\usepackage{appendix}
\usepackage{here}
\usepackage{url}
\usepackage{hyperref}
\usepackage{autobreak}
\usepackage{svg}
\usepackage[utf8]{inputenc}
\DeclareUnicodeCharacter{0306}{}

\usepackage{comment}

\renewcommand{\v}[1]{\boldsymbol{#1}}
\usepackage{tikz}
\usetikzlibrary{shapes,calc}
\usetikzlibrary{shapes.geometric,arrows.meta,decorations.markings}

\begin{document}

\title{Two-particle correlation effects on nonlinear optical responses in the 1d interacting Rice-Mele model}
\author{Akira Kofuji}
\email{kofuji.akira.46c@st.kyoto-u.ac.jp}
 \affiliation{Department of Physics, Kyoto University, Kyoto 606-8502, Japan}
\author{Robert Peters}%
 \email{peters@scphys.kyoto-u.ac.jp}
 \affiliation{Department of Physics, Kyoto University, Kyoto 606-8502, Japan}
\date{\today}

\begin{abstract}Nonlinear responses in crystalline solids are attracting a great deal of attention because of exciting phenomena, such as the bulk photovoltaic effect in noncentrosymmetric crystals and the third harmonic generation related to Higgs modes in superconductors, and their potential applicability to electronic devices. Recently,  nonlinear responses have also been studied in strongly correlated electron systems. Experimental evidence has revealed that correlations play a significant role in nonlinear responses. However, most theoretical calculations only consider excitonic effects or involve numerically demanding approaches, making interpreting the results challenging.
In this paper, we adopt another approach, which is based on real-time evolution using the correlation expansion method.
Particularly, we focus on the 1d interacting Rice-Mele model. We analyze the impact of the density-density interaction on the linear and nonlinear conductivities and demonstrate that two-particle correlations beyond the mean-field level enhance second-order nonlinear responses, especially the second harmonic generation, while the linear response is not strongly affected. Furthermore, by decomposing the current into a one-particle contribution and six two-particle contributions, we show that the ``biexciton transition" term and its nonlinear oscillations is the most dominant two-particle contribution to the nonlinear response. In addition, we also show that the intercell charge-charge correlation is strongly enhanced when the system is driven with the frequency corresponding to the excitonic peak and can even exceed the intracell correlation. This implies the possibility of manipulating two-particle correlations with external fields.
\end{abstract}

\maketitle

\section{Introduction}
\label{intro}
Optical and transport measurements are one of the most fundamental probes to study the microscopic properties of materials in condensed matter.
The linear response theory developed by R. Kubo successfully describes many optical responses and transport phenomena\cite{doi:10.1143/JPSJ.12.570}, making it possible to study not only the density of states but also, e.g., the topological nature of Bloch wave functions when probing the integer quantum Hall effect\cite{TKNN}. 
For stronger external fields or when the linear response is, e.g., prohibited by symmetries, the system's response is nonlinear.
Nonlinear responses are also related to microscopic material properties. For example, two-photon absorption processes make it possible to access one-photon forbidden states, the nonlinear Hall effect is related to the Berry curvature dipole\cite{Sodemann_2015}, which is the gradient of the Berry curvature in the momentum space, the Shift current is caused by the difference between the Berry connections of the conduction band and the valence band\cite{Baltz_1981,Sipe_2000,morimoto2016topological}, and third harmonic generation can detect a Higgs mode in superconductors\cite{Tsuji_2015, shimano_review}. These probes are unique to nonlinear responses, and thus, they are not only interesting for understanding materials but also essential for future applications to electronic devices such as solar cells and ultrafast optical switches\cite{solar_cell_review,2d_nonlinear_optical_review}.

Recently, nonlinear responses have also been studied in strongly correlated electron systems. Because of the intricately intertwined degrees of freedom in these systems, i.e., the coupling of charge, spin, and orbital, nonlinear responses become increasingly diverse. 
A gigantic optical nonlinearity in one-dimensional (1d) Mott insulators\cite{kishida2000gigantic,kishida_2001}, nonreciprocal transport originating in spin fluctuations in the chiral magnet MnSi\cite{yokouchi2017electrical}, a giant spontaneous Hall effect in the Weyl-Kondo semimetal candidate Ce$_3$Bi$_4$Pd$_3$\cite{Dzsaber_2021}, and an exciton-mediated enhancement of second harmonic generation(SHG) in 2d materials have been observed in experiments\cite{shree2021interlayer, wang2021giant}. These experiments demonstrate the importance and impact of strong Coulomb interaction on nonlinear responses. Furthermore, interest is growing in realizing thermodynamically inaccessible states by controlling non-equilibrium steady states and fluctuations by intense light irradiation\cite{Fausti_2011,KOSHIHARA20221}, where pump-probe spectroscopy is often utilized as a powerful tool\cite{chemla2001many,Basov_2011}. Thus, investigating nonlinear responses in correlated systems, preferably using a real-time approach, is highly desired.

Correlation effects on nonlinear responses have been studied theoretically, often paying particular attention to electron-hole interactions in semiconductors\cite{Takagahara_1986,Wang_1993,Stahl_1994,maialle_1994,madarasaz_1994,Ostreich_1995,Ostreich_1998,Kwong_2000,Ostreich_2001,takayama2002t,Turkowski_2014,mads_2014,Attaccalite_2017,Taghizadeh_2018,Morimoto_2020,Fei_2020,Kaneko_2021,Konabe_2021,Chan_2021,ruan2023excitonic,yutzu_2023}. 
In theoretical calculations, it has been shown that excitonic effects can significantly enhance the shift current in monolayer GeS\cite{Chan_2021}.
In other calculations, the enhancement of nonlinear responses related to the electron mass renormalization in heavy fermion systems\cite{Michishita_2021,Kofuji_2021}, spin-charge separation in 1d Mott insulators\cite{mizuno2000nonlinear}, and the interplay between charge transfer and electron correlations in charge-transfer Mott insulators\cite{zhang_2001} have been revealed by utilizing dynamical mean field theory and exact diagonalization.
Although our understanding of nonlinear responses in correlated systems grows, numerous open problems remain. 
Many previous approaches have been limited to excitonic effects on nonlinear responses.
Other approaches rely on numerically expensive techniques, such as the density matrix renormalization group.
Therefore, another approach to nonlinear responses is required that is not limited to particular degrees of freedom, can include correlation effects, and is easy to interpret.

In this paper, we use another numerical approach to analyze correlation effects on nonlinear optical responses, examining the impact of two-particle correlation effects. Our approach is based on the equation of motion using the correlation expansion\cite{FRICKE1996479}, including two-particle correlations but neglecting three-particle correlations. By comparing different approximations, we confirm the accuracy of this approach for weak to moderate interaction strengths.
Particularly, we consider the 1d interacting Rice-Mele model irradiated by an AC electric field and calculate the bulk photovoltaic effect, the SHG, and the time evolution of charge correlations in real space, taking the dynamics of two-particle correlations into account. We show that two-particle correlation effects are significant for nonlinear responses, while the linear response is not strongly affected except for a slight shift of the spectrum. Both the bulk photovoltaic effect and the SHG are enhanced by two-particle correlations, especially near the excitonic peak, and correlation effects are salient for the SHG. Moreover, we decompose the current into various contributions and show that a large part of the two-particle correlation effects arises from the term called ``biexciton transition"\cite{Stahl_1994} correlation. 
 Our approach can be easily extended to other systems, is based on a real-time approach compatible with pump-probe experiments, and can be used as a guideline to choose meaningful interactions when using a perturbation expansion. It can complement other 
 approaches, such as exact diagonalization, time-dependent DMRG, and non-equilibrium DMFT.

The rest of this paper is structured as follows: In Sec.~\ref{model_and_methods}, we describe our model and methods. Section~\ref{results_section_conductivity} shows the calculated conductivities of the linear absorption, the photovoltaic effect, and the SHG. In Sec.~\ref{decomposition_of_SHG}, we show the decomposed SHG to analyze the importance of two-particle correlations.
In Sec.~\ref{results_nonlinearity}, we demonstrate that  
two-particle correlations are essential for understanding the SHG conductivity. In Sec.~\ref{results_enhancement_correlations}, we show that the electric field enhances the short-range charge correlations of the system. Finally, in Sec.~\ref{conclusion}, we conclude and summarize the paper.

\section{Model \& Methods}
\label{model_and_methods}
\subsection{Hamiltonian and Current operator}
In this paper, we consider the 1d interacting spinless Rice-Mele model, a minimal inversion symmetry-broken model with nearest neighbor interaction. The effect of an electric field is incorporated by the Peierls phase. The Hamiltonian in real space is written as
\begin{equation}
\begin{aligned}
\hat{H}(t) &= \hat{H}_{0}(t)+\hat{H}_{int},\\ \hat{H}_{0}(t) &= \frac{Q_{x}-Q_{y}}{2}\sum_{i} (e^{-iA(t)/2} c_{i,A}^{\dagger} c_{i,B}+h.c.) \\&+ \frac{Q_{x}+Q_{y}}{2}\sum_{i} (e^{-iA(t)/2} c_{i,B}^{\dagger} c_{i+1,A}+h.c.) \\ &+ Q_{on}\sum_{i} (n_{i,A}-n_{i,B}) ,\\\hat{H}_{int} &=  V\sum_{i} n_{i,A}(n_{i,B}+n_{i+1,B}),
\end{aligned}
\end{equation}
where $c_{i,a}^{\dagger}$($c_{i,a}$) are creation(annihilation) operators for electrons on the sublattice a = $\{A,B\}$. $n_{i,a}=c_{i,a}^{\dagger}c_{i,a}$ is the occupation operator for sublattice a, and $A(t)$ is the external vector potential. 
$Q_x$, $Q_y$, and $Q_{on}$ correspond to hopping amplitudes and the local potential. $V$ is the strength of the nearest-neighbor density-density interaction.
Throughout this paper, we set the Planck constant, the lattice constant, and the electron charge to unity, $\hbar=a=e=1$. Other parameters are fixed to $Q_{x}=0.25$, $Q_{y}=0.3$, and $Q_{on}=0.25$. The strength of the interaction is varied between $V=0.0,\ldots, 0.15$.

The current operator $\hat{J}(t)$ is defined as the derivative of the Hamiltonian with respect to the vector potential:
\begin{equation}
\begin{aligned}
\hat{J}(t) &= -\frac{\partial \hat{H}(t)}{\partial A(t)} \\
&=i\frac{Q_{x}-Q_{y}}{4}\sum_{i} (e^{-iA(t)/2} c_{i,A}^{\dagger} c_{i,B}-h.c.)\\ &+ i\frac{Q_{x}+Q_{y}}{4}\sum_{i} (e^{-iA(t)/2} c_{i,B}^{\dagger} c_{i+1,A}-h.c.).
\end{aligned}
\end{equation}
Next, we Fourier transform this model to the momentum space and introduce the Houston basis\cite{Krieger,murakami2022doping,Nagai_2023}. 
We define the Fourier transform of $c_{i,A/B}^{\dagger}$ to $c_{k,A/B}^{\dagger}$ as:
\begin{equation}
\begin{aligned}
c_{k,A/B}^{\dagger} = \frac{1}{\sqrt{N}}\sum_{i}e^{ikr_{i}}c_{i,A/B}.
\end{aligned}
\end{equation}
$r_{i}$ is the position of the $i$-th unit cell, and $N$ is the number of unit cells in the system. In this definition, the position in the unit cell is not included.
The non-interacting part of the Hamiltonian can be written as 
\begin{equation}
\hat{H}_{0}(t)=\sum_{k}[c_{k,A}^{\dagger} c_{k,B}^{\dagger}]W(t)^{\dagger}H(k_t)W(t)[c_{k,A} c_{k,B}]^{T},
\end{equation}
where $k_t=k-A(t)$, and $W(t)$ is defined as
\begin{equation}
\begin{aligned}
W(t) = 
\begin{bmatrix}
1 && 0 \\
0 && e^{iA(t)/2} \\
\end{bmatrix}.
\end{aligned}
\end{equation}
$H(k_{t})$ can be diagonalized by a unitary matrix $U(k_{t})$ at each time step.  The eigenenergies of $H_{0}(k_{t})$ are $\varepsilon_{c/v} (k_{t})=\pm \sqrt{Q_{x}^{2}\cos^{2} (k_{t}/2) + Q_{y}^{2}\sin^{2} (k_{t}/2) + Q_{on}^{2}}$. We use the eigenstates of the diagonalized non-interacting part of the Hamiltonian in momentum space as a new basis, commonly called the Houston basis, describing the valence (v) and conduction (c) bands. 
The Houston basis is defined as follows:
\begin{equation}
\begin{aligned}
\begin{bmatrix}
c_{k,c} \\
c_{k,v}
\end{bmatrix}
&= U(k_{t})^{\dagger} W(t)^{\dagger}
\begin{bmatrix}
c_{k,A} \\
c_{k,B}
\end{bmatrix},\\
U(k_{t})&=
\begin{bmatrix}
U_{Ac}(k_{t}) && U_{Av}(k_{t})\\
U_{Bc}(k_{t}) && U_{Bv}(k_{t})
\end{bmatrix}.
\end{aligned}
\end{equation}
We note that the Houston basis depends on the time when the vector potential is finite because of the time-dependent unitary transformation.
Hereafter, we denote the matrix representation of an operator in the Houston basis using a tilde, i.e., $\hat{O}=\sum_{k}[c_{k,c}^{\dagger} c_{k,v}^{\dagger}]\tilde{O}(k_{t})[c_{k,c} c_{k,v}]^{T}$. 
 The non-interacting Hamiltonian, the dipole matrix, and the current operator in the Houston basis are as follows:
\begin{equation}
\begin{aligned}
\tilde{H}(k_{t}) &=
\begin{bmatrix}
\varepsilon_{c} (k_{t}) && 0 \\
0 && \varepsilon_{v} (k_{t})
\end{bmatrix}
-\tilde{d}(k_{t})E(t),\\
\tilde{d}(k_{t})&=iU(k_{t})^{\dagger}V(k)^{\dagger}\frac{\partial}{\partial k}(V(k)U(k_{t}))\\
&=
\begin{bmatrix}
d_{cc}(k_{t}) && d_{cv}(k_{t}) \\
d_{vc}(k_{t}) && d_{vv}(k_{t})
\end{bmatrix},\\
\tilde{J}(k_{t}) &= 
\begin{bmatrix}
\frac{\partial \varepsilon_{c}(k_{t})}{\partial k} && 2i\varepsilon_{c}(k_{t}) d_{cv}(k_{t})\\
-2i\varepsilon_{c}(k_{t}) d_{vc}(k_{t}) && \frac{\partial \varepsilon_{v}(k_{t})}{\partial k}
\end{bmatrix},
\end{aligned}
\end{equation}
where $V(k)$ is a unitary matrix defined as:
\begin{equation}
\begin{aligned}
V(k) = 
\begin{bmatrix}
1 && 0 \\
0 && e^{-ik/2}
\end{bmatrix}.
\end{aligned}
\end{equation}
$V(k)$ arises from the fact that our definition of the Fourier transform does not include the position inside the unit cell.
$E(t)=-\frac{\partial A(t)}{\partial t}$ is the external electric field.
The off-diagonal terms in the Hamiltonian proportional to the dipole matrix arise from the time derivative of the unitary matrix because the Houston basis is time-dependent.
Finally, the interacting part of the Hamiltonian $\hat{H}_{\mathrm{int}}$ in the Houston basis is given as
\begin{equation}
\begin{aligned}
&\hat{H}_{\mathrm{int}} = \sum_{\substack{k,k^\prime,q \\ \alpha,\beta,\gamma,\delta}}f_{\alpha\beta\gamma\delta}(k_{t},k_{t}^\prime,q) c_{k+q,\alpha}^{\dagger}c_{k^\prime-q,\beta}^{\dagger}c_{k^\prime,\gamma}c_{k,\delta},\\
&f_{\alpha \beta \gamma \delta}(k_{t},k_{t}^\prime,q)=\frac{V}{N}(1+e^{-iq})U_{A\alpha}^{*}({k_{t}+q}) U_{B\beta}^{*}({k_{t}^\prime-q})\\&~~~~~~~~~~~~~~~~~~~~~~~~~~\times U_{B\gamma}({k_{t}^\prime}) U_{A\delta}({k_{t}}),
\end{aligned}
\end{equation}
where $\alpha,\beta,\gamma,\delta=\{c,v\}$, corresponding to the two basis states of the Houston basis.
Using fermionic commutation relations, several terms can be combined. For example,
terms proportional to $c_{k+q,c}^{\dagger}c_{k^{'}-q,c}^{\dagger}c_{k^{'},v}c_{k,c}$ can be combined with terms $c_{k+q,c}^{\dagger}c_{k^{'}-q,c}^{\dagger}c_{k^{'},c}c_{k,v}$. Thus, there are $9$ types of interactions in the Houston basis, corresponding to $C^{\dagger}C^{\dagger}CC$, $C^{\dagger}C^{\dagger}CV$, $C^{\dagger}C^{\dagger}VV$, $C^{\dagger}V^{\dagger}CC$, $C^{\dagger}V^{\dagger}CV$, $C^{\dagger}V^{\dagger}VV$, $V^{\dagger}V^{\dagger}CC$, $V^{\dagger}V^{\dagger}CV$, and $V^{\dagger}V^{\dagger}VV$. Defining the coefficients of these terms as $F_{\alpha\beta\gamma\delta}(k,k^{'},q)$, we can write the interacting part of the Hamiltonian as
\begin{equation}
\begin{aligned}
&\hat{H}_{\mathrm{int}} = \sum_{\substack{k,k^{'},q \\ \alpha,\beta,\gamma,\delta}}F_{\alpha\beta\gamma\delta}(k_{t},k_{t}^{'},q) c_{k+q,\alpha}^{\dagger}c_{k^{'}-q,\beta}^{\dagger}c_{k^{'},\gamma}c_{k,\delta}\\
\end{aligned}
\end{equation}
where $(\alpha,\beta,\gamma,\delta)=\{(c,c,c,c)$, $(c,c,c,v)$, $(c,c,v,v)$, $(c,v,c,c)$, $(c,v,c,v)$, $(c,v,v,v)$, $(v,v,c,c)$, $(v,v,c,v)$, $(v,v,v,v)\}$.

\subsection{Equation of motion}
\subsubsection{Correlation expansion}
We utilize the correlation expansion method by Fricke\cite{FRICKE1996479} to calculate the time evolution of this system. This method is useful as a closed set of equations of motion(EOM) for correlation functions can be obtained systematically. Here, we briefly review this method. For more information, we refer to the original paper\cite{FRICKE1996479}.

The Heisenberg EOM for an operator is
\begin{equation}
    \dot{\hat{O}} =-i [\hat{O},\hat{H}(t)].
\end{equation}
If the Hamiltonian includes two-particle operators, e.g., $V \sum_{i} n_{i,A}(n_{i,B}+n_{i+1,B})$ in our case, the commutator between a one-particle operator and $\hat{H}(t)$ yields two-particle operators, the commutator between a two-particle operator and $\hat{H}(t)$ yields three-particle operators, the commutator between a three-particle operator and $\hat{H}(t)$ yields four-particle operators, and so on. Thus, even if we are interested in one-particle quantities, we have to know all higher-order many-particle quantities, which is called the hierarchy problem. The correlation expansion method is a simple prescription to truncate this hierarchy. Each expectation value is expanded in correlations, denoted as $\langle .\rangle^C$.
We can write the correlation expansion symbolically as follows:
\begin{equation}
\label{correlation_expansion}
\begin{aligned}
\langle B_{1} \rangle &= \langle B_{1} \rangle^{c},\\
\langle B_{1}B_{2} \rangle &= \langle B_{1}B_{2} \rangle^{c}+\langle B_{1} \rangle^{c}\langle B_{2} \rangle^{c},\\
\langle B_{1}B_{2}B_{3} \rangle &= \langle B_{1}B_{2}B_{3} \rangle^{c} + \langle B_{1}B_{2} \rangle^{c}\langle B_{3} \rangle^{c}\\ &+ \langle B_{2}B_{3} \rangle^{c}\langle B_{1} \rangle^{c} + \langle B_{3}B_{1} \rangle^{c}\langle B_{2} \rangle^{c}\\ &+ \langle B_{1} \rangle^{c}\langle B_{2} \rangle^{c}\langle B_{3} \rangle^{c},
\end{aligned}
\end{equation}
where 
$B_{i}$ is an arbitrary product of creation and annihilation operators. For higher orders, the procedure is defined recursively. For clarity, we look at the following example: $\langle c_{k+q,c}^{\dagger} c_{k^{'}-q,c}^{\dagger} c_{k^{'},c} c_{k,v} \rangle$. The correlation expansion for this expectation value can be written as
\begin{equation}
\label{correlation_expansion_example}
\begin{aligned}
\langle c_{k+q,c}^{\dagger} c_{k^{'}-q,c}^{\dagger} c_{k^{'},c} c_{k,v} \rangle &= \langle c_{k+q,c}^{\dagger} c_{k^{'}-q,c}^{\dagger} c_{k^{'},c} c_{k,v} \rangle^{c} \\
&+ \langle c_{k+q,c}^{\dagger} c_{k,v} \rangle^{c} \langle c_{k^{'}-q,c}^{\dagger} c_{k^{'},c} \rangle^{c} \\
&- \langle c_{k+q,c}^{\dagger} c_{k^{'},c} \rangle^{c} \langle c_{k^{'}-q,c}^{\dagger} c_{k,v} \rangle^{c} \\
&= S_{cccv}(k,k^{'},q) + \delta_{q,0} y(k)f_{c}(k^{'}) \\
&- \delta_{k+q,k^{'}} f_{c}(k^{'}) y(k),
\end{aligned}
\end{equation}
where we define the following quantities:
\begin{equation}
\label{f_y_definition}
\begin{aligned}
S_{\alpha \beta \gamma \delta}(k,k^{'},q) &=\langle c_{k+q,\alpha}^{\dagger} c_{k^{'}-q,\beta}^{\dagger} c_{k^{'},\gamma} c_{k,\delta}\rangle^{c},\\ f_{\alpha}(k) &=\langle  c_{k,\alpha}^{\dagger}c_{k,\alpha}\rangle,\\ y(k) &= \langle c_{k,c}^{\dagger}c_{k,v} \rangle.
\end{aligned}
\end{equation}
In the correlation expansion, Eq.~(\ref{correlation_expansion}), a $n$-particle expectation value is decomposed into a  $n$-particle correlation function and terms, which can be written as a product of $m(<n)$-particle correlations. One-particle correlations are identical to the corresponding one-particle expectation values. For example, two-particle expectation values are decomposed into the product of one-particle expectation values and a two-particle correlation function. 
We note that a $n$-particle correlation ($n>1$) fulfills fermionic symmetries,
\begin{equation}
\begin{aligned}
\langle \cdots c_{k_{1},\alpha_{1}}^{\dagger} c_{k_{2},\alpha_{2}}^{\dagger} \cdots\rangle^{c} &= -\langle \cdots c_{k_{2},\alpha_{2}}^{\dagger} c_{k_{1},\alpha_{1}}^{\dagger} \cdots\rangle^{c},\\
\langle \cdots c_{k_{1},\alpha_{1}}^{\dagger} c_{k_{2},\alpha_{2}} \cdots\rangle^{c}&=-\langle \cdots c_{k_{2},\alpha_{2}} c_{k_{1},\alpha_{1}}^{\dagger} \cdots\rangle^{c}.
\end{aligned}
\end{equation}
We can rewrite the EOM using these correlation functions. The EOM can be expressed in the following way:
\begin{equation}
\label{EOM_before_truncate}
\begin{aligned}
\langle \dot{\mathrm{1P}} \rangle^{c} &=\langle \mathrm{1P} \rangle^{c} +  \langle \mathrm{1P} \rangle^{c} \langle \mathrm{1P} \rangle^{c}+ \langle \mathrm{2P} \rangle^{c}\\
\langle \dot{\mathrm{2P}} \rangle^{c} &=  \langle \mathrm{1P} \rangle^{c}\langle \mathrm{1P} \rangle^{c} +  \langle \mathrm{2P} \rangle^{c} \\
&+ \langle \mathrm{1P} \rangle^{c}\langle \mathrm{1P} \rangle^{c}\langle \mathrm{1P} \rangle^{c}+\langle \mathrm{1P} \rangle^{c}\langle \mathrm{2P} \rangle^{c} \\
&+\langle \mathrm{3P} \rangle^{c}.
\end{aligned}
\end{equation}
$\langle \mathrm{1P} \rangle^{c}$, $\langle \mathrm{2P} \rangle^{c}$, and $\langle \mathrm{3P} \rangle^{c}$ express one-particle terms, two-particle terms, and three-particle terms respectively. 
Equation~(\ref{EOM_before_truncate}) is not a closed set of equations. 
However, if we neglect the three-particle correlation terms, a closed set of EOM can be obtained, including two-particle correlations. 
This is motivated by the fact that all correlation functions involving more than one particle vanish in a non-interacting system.
Thus, if the interaction strength is not too strong, higher-order correlations are not essential, and this truncation can be expected to be appropriate. 
In this paper, we consider parameter regions where the interaction $V$ is moderate and the system is adiabatically connected to the non-interacting system, $V=0$. Therefore, we neglect three-particle and higher correlations.

The EOM arising from $\hat{H_{0}}(t)$ are
\begin{equation}
\label{eq_EOM_H0}
\begin{aligned}
\dot{f_{c}}(k)|_{H_{0}}&= -2\mathrm{Im}(d_{cv}(k_{t})E(t)y(k))-\gamma(f_{c}(k)-f_{c0}(k)),\\
\dot{f_{v}}(k)|_{H_{0}}&= 2\mathrm{Im}(d_{cv}(k_{t})E(t)y(k))-\gamma(f_{v}(k)-f_{v0}(k)),\\
\dot{y}(k)|_{H_{0}}&= -i(-2\varepsilon_{c}(k_{t})+E(t)(d_{cc}(k_{t})-d_{vv}(k_{t})))y(k)
\\&-iE(t)d_{cv}^{*}(k_{t})(f_{v}(k)-f_{c}(k)) -\gamma(y(k)-y_{0}(k)),
\end{aligned}
\end{equation}
where we have phenomenologically introduced relaxation terms originating in a coupling of the system to other degrees of freedom that are not included in our formalism, e.g., phonons. $\gamma$ is the relaxation rate and $f_{c0}(k)$, $f_{v0}(k)$, and $y_{0}(k)$ are the equilibrium values of $f_{c}(k)$, $f_{v}(k)$, and $y(k)$ in the ground state, as defined in Eq.~(\ref{f_y_definition}). Such relaxation terms are also included in the EOM for the two-particle correlations. They drive the system back to the equilibrium state. $\gamma$ is fixed to $\gamma=0.02$ throughout this paper.

\subsubsection{Time-dependent mean-field equations}
To analyze the importance of two-particle correlations, we will furthermore compare the results to a set of equations where two-particle correlations are also neglected. This corresponds to the time-dependent mean-field approximation (tdMF).
Besides Eq.~(\ref{eq_EOM_H0}), the EOM includes one-particle correlations originating in $H_{\mathrm{int}}$ as 
\begin{equation}
\label{MF_EOM}
\begin{aligned}
\dot{f_{c}}(k)|_{\mathrm{MF}}&= 2\mathrm{Im}(M_{cv}(k)y(k)),\\
\dot{f_{v}}(k)|_{\mathrm{MF}}&= -2\mathrm{Im}(M_{cv}(k)y(k)),\\
\dot{y}(k)|_{\mathrm{MF}}&= -i(M_{vv}(k)-M_{cc}(k))y(k)
\\&+iM_{vc}(k)(f_{v}(k)-f_{c}(k)).
\end{aligned}
\end{equation}
$M_{\alpha\beta}(k)$ are mean-field corrections to the Hamiltonian. Their explicit expressions are given in the Appendix~\ref{M_definition}.

\subsubsection{Initial state}
Finally, we describe the initial state used for calculating the time evolution. 
In principle, we use the ground state as the initial state for each calculation. For tdMF calculation, the ground state can be obtained by diagonalizing the mean-field Hamiltonian. On the other hand, a direct diagonalization of the Hamiltonian is not possible for calculations including two-particle correlations. In this situation, the ground state is obtained by adiabatically switching on the interaction, which is possible for the weak to moderate interaction strengths used in our calculation. More details about this procedure are explained in the Appendix~\ref{appendix_ground_state}

\subsection{Conductivity}
\label{cond}
\begin{figure*}[t]
\begin{center}
\includegraphics[width=0.32\linewidth]{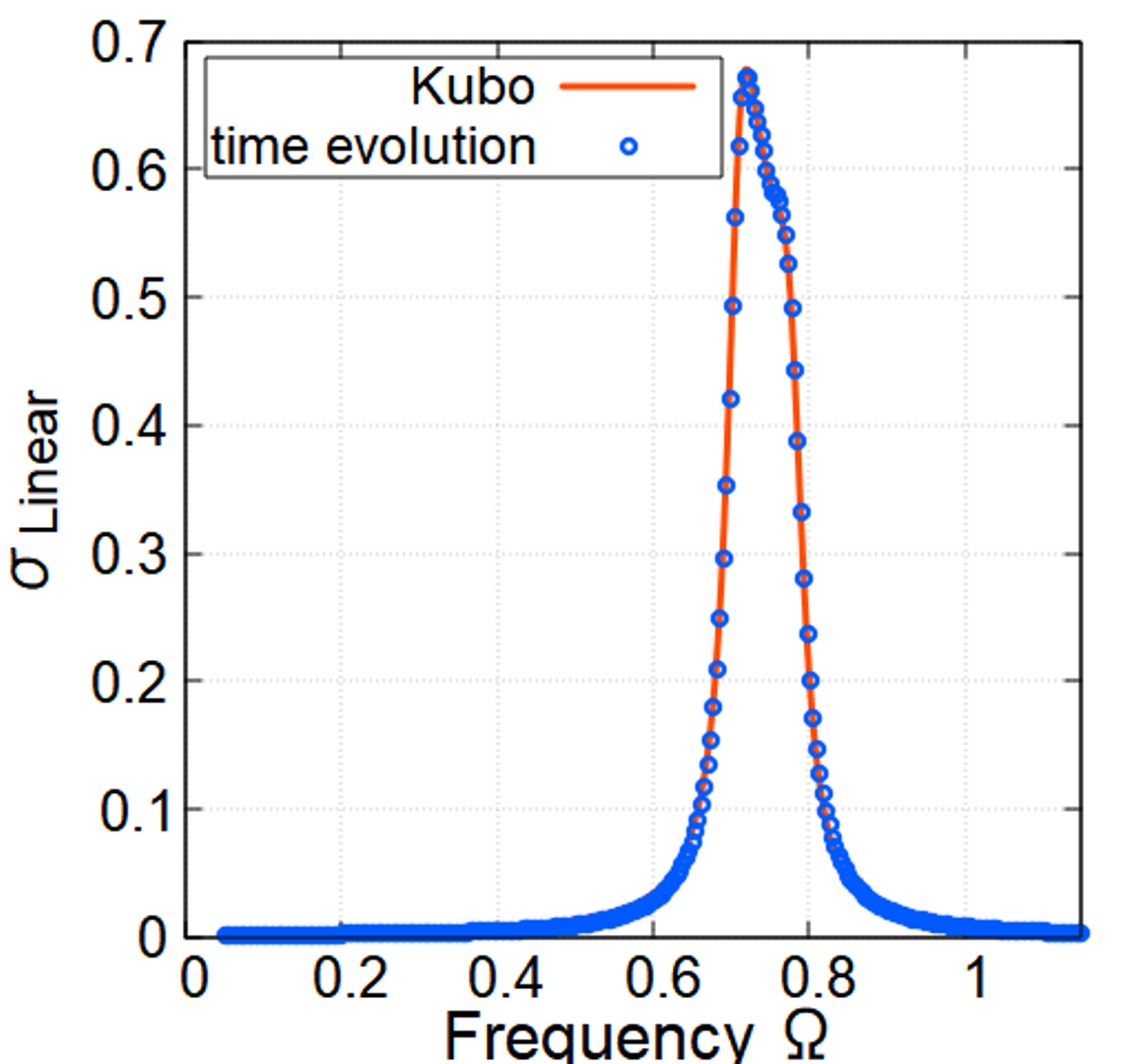}
\includegraphics[width=0.32\linewidth]{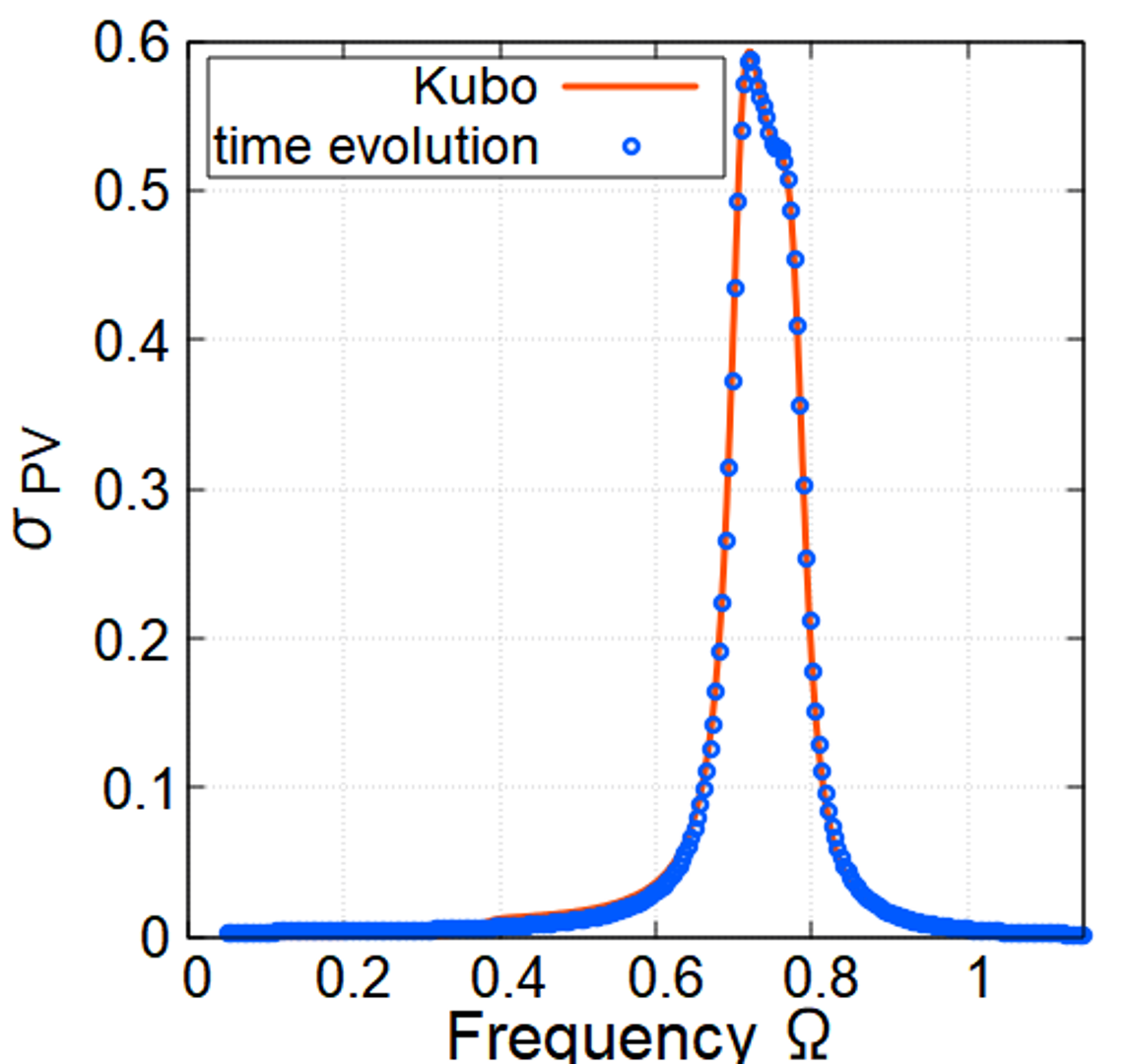}
\includegraphics[width=0.32\linewidth]{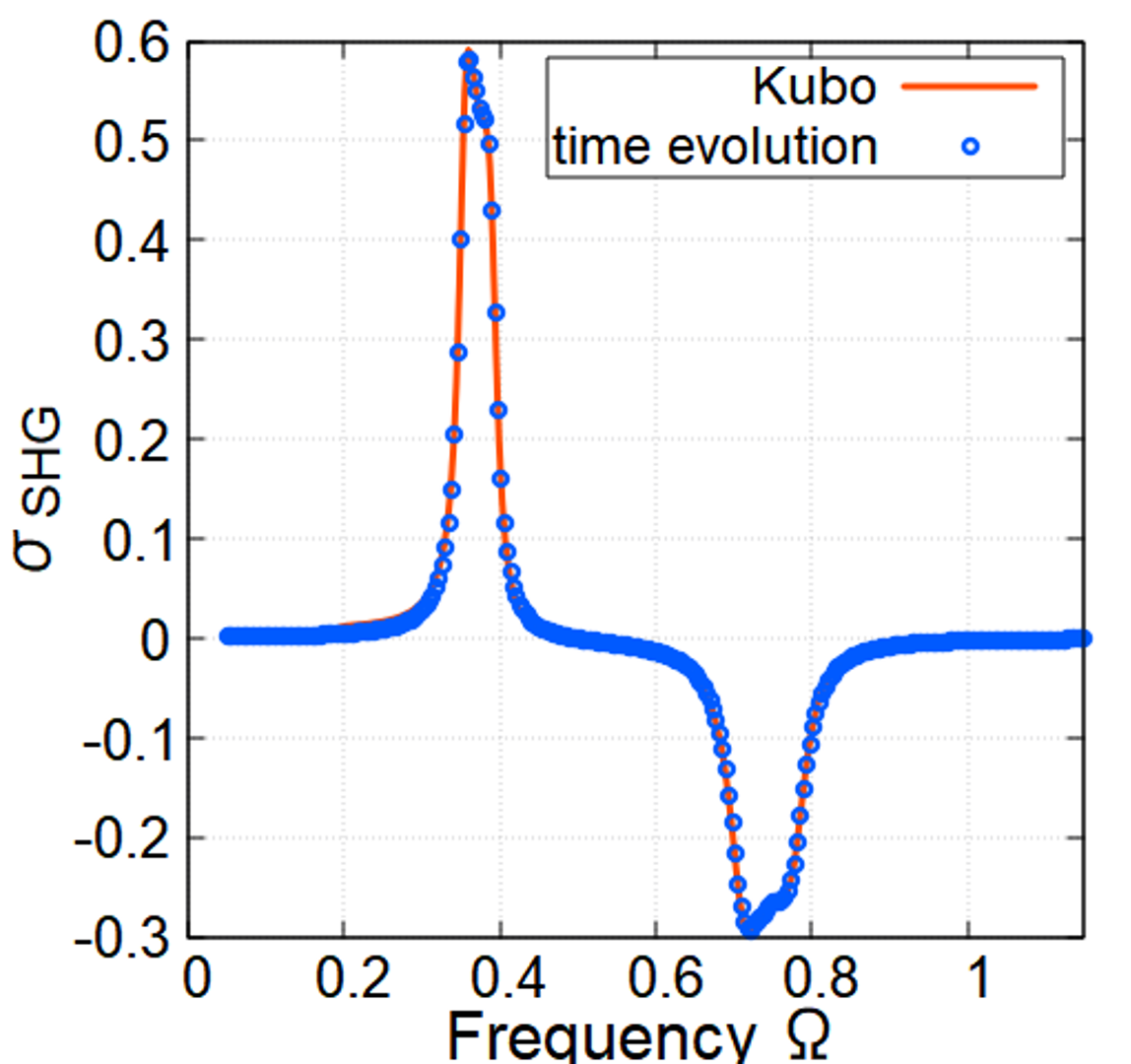}
\end{center}
\caption{Comparison between conductivities calculated by the Kubo formula(Kubo) and tdMF(tdMF) for $V=0$. Left panel: linear conductivity $\sigma_{\mathrm{Linear}}$. Middle panel: photovoltaic conductivity $\sigma_{\mathrm{PV}}$. Right panel: SHG conductivity $\sigma_{\mathrm{SHG}}$.}
\label{conductivity_check}
\end{figure*}
There are two main approaches to calculating the conductivity of a system. One is an analytical approach using the Kubo formula and its extensions to nonlinear responses. The other approach is to calculate the time evolution of the system. Recently, calculating the time evolution has been adopted more frequently because it can easily take temporal fluctuations into account and is more compatible with pump-probe experiments. Here, we use the time-evolution approach to calculate the linear, photovoltaic, and SHG conductivities.

We briefly explain how to calculate conductivities in this approach. First, we calculate the time evolution of the system until we obtain a non-equilibrium steady state. We note that relaxation terms play an essential role in stabilizing a non-equilibrium steady state. 
Then, the current at each time step can be calculated by one-particle quantities, $f_{c/v}(k)$ and $y(k)$, as
\begin{equation}
\begin{aligned}
\langle \hat{J} \rangle = -\frac{\partial \varepsilon_{c}(k_{t})}{\partial k}(f_{v}(k)-f_{c}(k))-4\varepsilon_{c}(k_{t})\mathrm{Im}[d_{cv}(k_{t})y(k)].
\end{aligned}
\end{equation}
If we consider an electric field with a single frequency, i.e., $E(t)=E_{0}\cos(\Omega t)$, the steady state and physical observables are periodic with period $\frac{2 \pi}{\Omega}$. Thus, we Fourier transform the expectation value of the current operator as
\begin{equation}
J_{n\Omega} = \int_{t_{0}}^{t_{0}+\Omega/2\pi} d\omega \langle \hat{J} \rangle e^{in\Omega t}.
\end{equation}
For weak electric fields $E_{0}$, $J_{1\Omega}$ corresponds to the linear response, and $J_{0\Omega}$ and $J_{2\Omega}$ correspond to second-order nonlinear responses (For strong electric fields, higher-order nonlinear responses will also affect these currents). In this paper, the electric field has the form $E(t)=E_{0}\cos(\Omega t)$, and the strength is fixed to $E_{0}=-0.005$ when calculating currents.
By dividing these conductivities by $E_{0}/2$ or $(E_{0}/2)^{2}$, we define the linear, photovoltaic, and SHG conductivities as
\begin{equation}
\begin{aligned}
\sigma_{\mathrm{Linear}} &= \mathrm{Re}\left[\frac{J_{1\Omega}}{E_{0}/2} \right],\\
\sigma_{\mathrm{PV}} &= \mathrm{Re}\left[\frac{J_{0\Omega}}{(E_{0}/2)^{2}} \right],\\
\sigma_{\mathrm{SHG}} &= \mathrm{Re}\left[\frac{J_{2\Omega}}{(E_{0}/2)^{2}} \right].
\end{aligned}
\end{equation}
To confirm that these conductivities are well-defined in our approach, we compare them with conductivities calculated by the Kubo formula in the non-interacting system. Figure~\ref{conductivity_check} shows the linear, photovoltaic, and SHG conductivities calculated by the Kubo formula and the EOM using the correlation expansion.
In all three cases, the conductivities calculated by both methods agree very well.
Tiny differences can be explained by the broadening, i.e., the relaxation,  which is introduced differently in the Kubo formula and the time-evolution formalism. These results demonstrate that the conductivities calculated with our approach are well-defined physical quantities.

To analyze the impact of correlations on the conductivity, we will compare three different levels of approximations.
Besides calculating the conductivity via the time evolution, including two-particle correlations (2P) and using time-dependent mean-field equations (tdMF), we will calculate the conductivity using the independent particle approximation (IPA). In the IPA, we calculate the ground state of the interacting system in equilibrium by neglecting two-particle correlations. Thus, interactions only result in renormalized parameters. (The expressions of the renormalized parameters are given in the appendix in Eq.~\ref{renormalization}.)
We then use the Kubo formula to calculate the conductivity in this system with renormalized parameters directly. Two-particle correlations, such as vertex corrections and the time dependence of expectation values, are entirely neglected in this approximation.

\subsection{Decomposition of the current}
\label{method_decomposition}
In our approach, there are various two-particle correlations, such as $S_{cccc}$, $ S_{cccv}$, $S_{ccvv}$, \dots $S_{vvvv}$, as defined in Eq.~(\ref{f_y_definition}). To clarify which two-particle correlations strongly affect the current, we decompose the current into various contributions. 

Although the current itself is a one-particle quantity and includes only $f_{c/v}(k)$ and $y_{k}$, the time derivative of the current includes various types of two-particle correlations. It can be symbolically written as (neglecting coefficients):
\begin{equation}
\begin{aligned}
\langle \dot{\hat{J}} \rangle &= \langle \mathrm{1P}\rangle^{c} + \langle \mathrm{1P}\rangle^{c} \langle \mathrm{1P}\rangle^{c} \\
&+ \langle C^{\dagger} C^{\dagger} C C\rangle^{c}\\
&+ \langle C^{\dagger} C^{\dagger} C V\rangle^{c}\\
&+ \langle C^{\dagger} C^{\dagger} V V\rangle^{c}\\
&+ \langle C^{\dagger} V^{\dagger} C V\rangle^{c}\\
&+ \langle C^{\dagger} V^{\dagger} V V\rangle^{c}\\
&+ \langle V^{\dagger} V^{\dagger} V V\rangle^{c}.
\end{aligned}
\end{equation}
$\langle \alpha^{\dagger} \beta^{\dagger} \gamma \delta \rangle^{c}$ corresponds to two-particle correlation terms related to $S_{\alpha \beta \gamma \delta}(k,k^{'},q)$.
Thus, the time derivative of the current can be decomposed into one-particle terms and six two-particle correlation terms.  Here, we combine $\langle V^{\dagger} V^{\dagger} C C \rangle^{c}$ with $\langle C^{\dagger} C^{\dagger} V V \rangle^{c}$ because they are related to each other by complex conjugation.
Similarly, we note that $\langle C^{\dagger} V^{\dagger} C C \rangle^{c}$ and $\langle C^{\dagger} C^{\dagger} C V \rangle^{c}$, and $\langle V^{\dagger} V^{\dagger} C V \rangle^{c}$ and $\langle C^{\dagger} V^{\dagger} V V \rangle^{c}$ can be combined respectively. 

Using this decomposition of the time derivative of the current, we can analyze the importance of two-particle correlations in the current.
We integrate the derivative to obtain the current as
\begin{equation}
\label{decomposed_current_int}
\begin{aligned}
\langle \hat{J} \rangle =  \int^{t}\langle \dot{\hat{J}} \rangle &= \int^{t}\langle \mathrm{1P}\rangle^{c} + \int^{t}\langle \mathrm{1P}\rangle^{c} \langle \mathrm{1P}\rangle^{c} \\
&+ \int^{t}\langle C^{\dagger} C^{\dagger} C C\rangle^{c}\\
&+ \int^{t}\langle C^{\dagger} C^{\dagger} C V\rangle^{c}\\
&+ \int^{t}\langle C^{\dagger} C^{\dagger} V V\rangle^{c}\\
&+ \int^{t}\langle C^{\dagger} V^{\dagger} C V\rangle^{c}\\
&+ \int^{t}\langle C^{\dagger} V^{\dagger} V V\rangle^{c}\\
&+ \int^{t}\langle V^{\dagger} V^{\dagger} V V\rangle^{c}.
\end{aligned}
\end{equation}
The current is decomposed into one-particle terms and six terms related to two-particle correlations. 

\section{Results}
\subsection{Linear and nonlinear conductivities}
\label{results_section_conductivity}
In this section, we analyze the linear, photovoltaic, and SHG conductivities for different interaction strengths, $V=0.0 \ldots 0.15$. We compare 
the results of one-particle mean-field calculations (tdMF) with the calculations, including two-particle correlations (2P), and results based on the IPA. As noted above, the parameters in the non-interacting Hamiltonian are $Q_{x}=0.25$, $Q_{y}=0.3$, and $Q_{on}=0.25$. The system consists of $22$ sites. The relaxation rate in the EOM is $\gamma=0.02$, and the strength of the electric field is $E_{0}=-0.005$.

\begin{figure}[t]
\begin{center}
\includegraphics[width=0.75\linewidth]{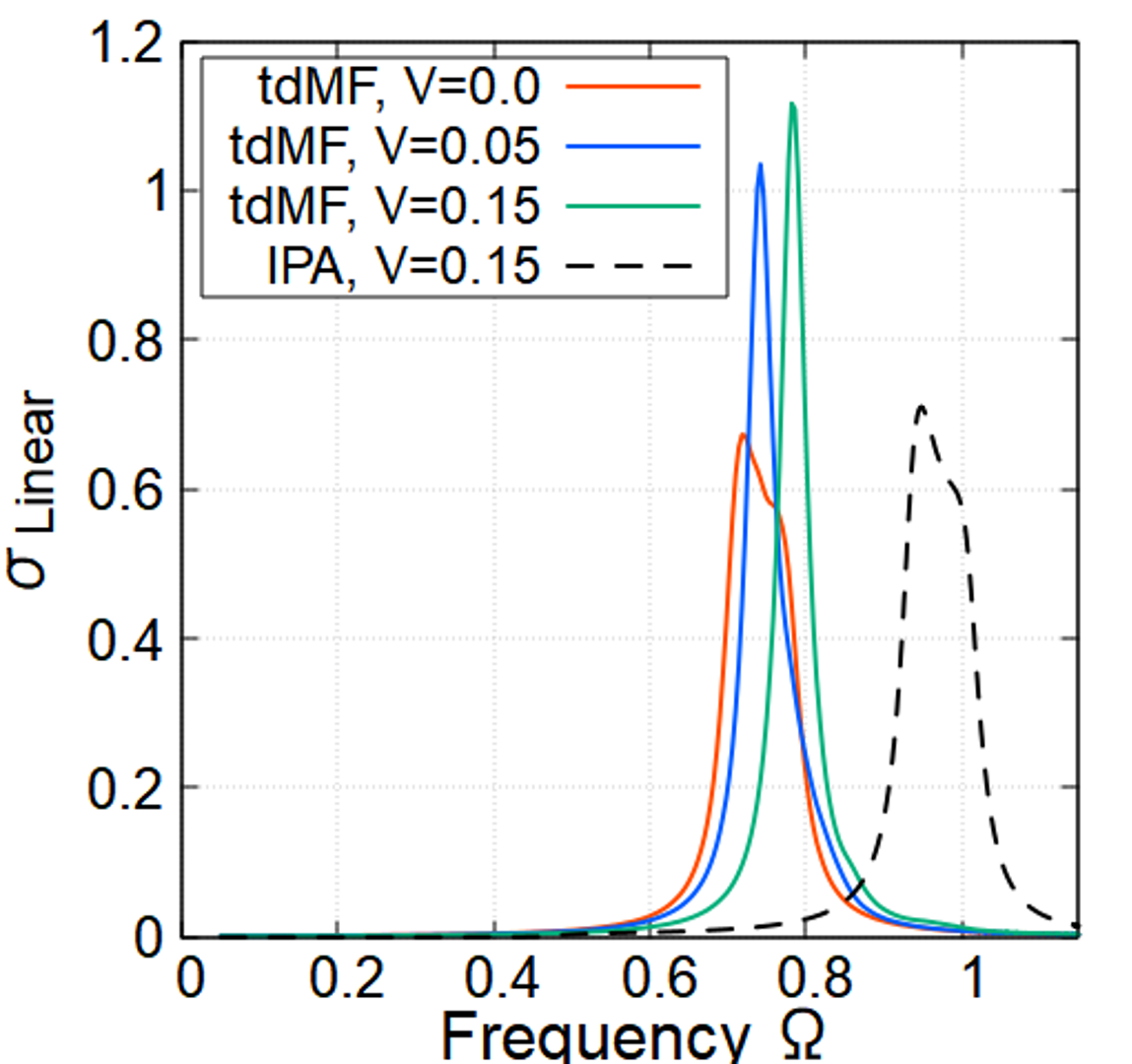}
\\
\includegraphics[width=0.75\linewidth]{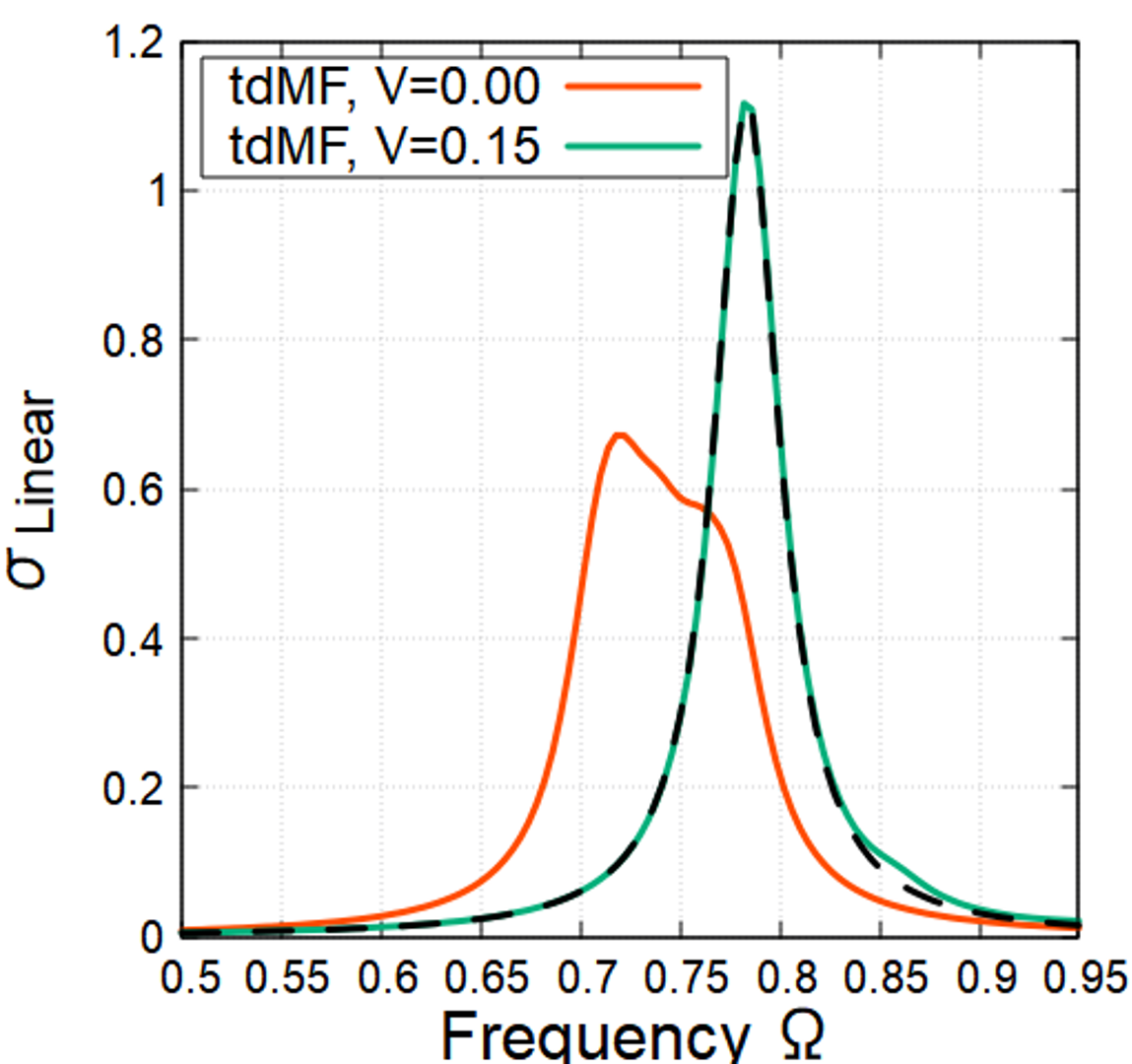}
\end{center}
\caption{Upper panel: linear conductivity calculated by tdMF for $V=0,0.05,0.15$ and linear conductivity calculated by IPA for $V=0.15$. Lower panel: the magnification of the upper panel, showing the linear conductivity around $\Omega = 0.7 \sim 0.8$ calculated by tdMF for $V=0.0,0.15$. The dashed line in the lower panel is a Lorentzian fit of the peak of the conductivity for $V=0.15$.}
\label{linear_cond_exciton}
\end{figure}

First, we focus on the linear conductivity. The upper panel of Fig.~\ref{linear_cond_exciton} shows the linear conductivity for $V=0.0,0.05,0.15$ calculated by tdMF. For comparison, we include the conductivity calculated by IPA for $V=0.15$.
As IPA only takes into account equilibrium expectation values and neglects dynamical fluctuations, the interaction between excited electrons and holes is not considered. Thus, IPA strongly overestimates the gap between the valence and conduction band and thus puts the spectrum at far too high frequencies.
All spectra calculated by tdMF are located at lower frequencies than the spectrum calculated by IPA. In Appendix~\ref{appendix_electron_hole_correlation}, we show that the peaks calculated by tdMF in the interacting system have an excitonic nature by analyzing the correlation function of the electron density and the hole density in real space. IPA fails to capture the excitonic nature of this peak correctly.
As the interaction is increased, the peak around $\Omega = 0.7 \sim 0.8$ becomes taller, and the width becomes smaller. The peak itself can be fitted by a Lorentzian function, which is demonstrated in the lower panel of Fig.~\ref{linear_cond_exciton} for $V=0.15$. 
We thus see that interactions enhance the response at the excitonic peak.

\begin{figure}[t]
\begin{center}
\includegraphics[width=0.73\linewidth]{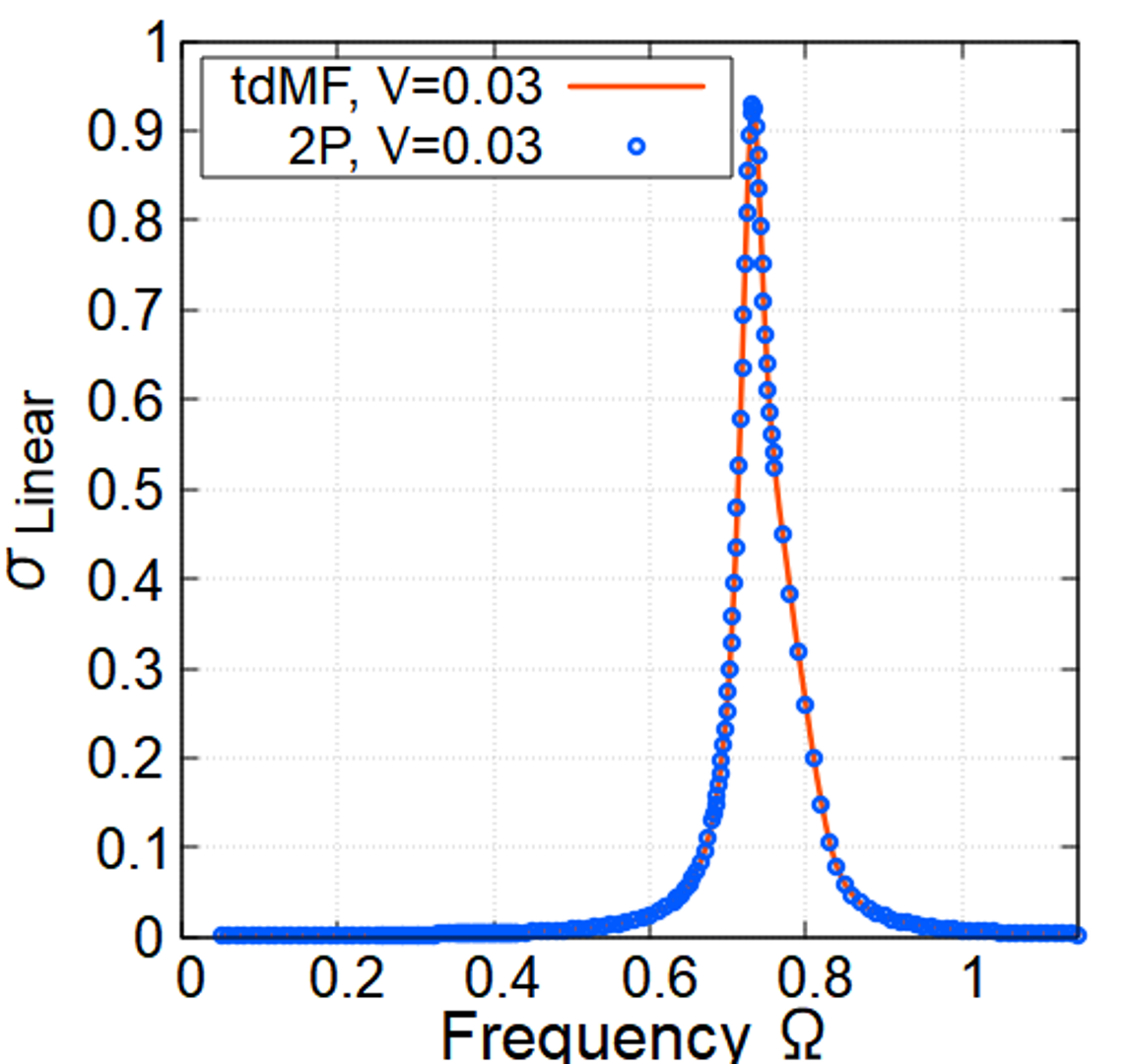}
\\
\includegraphics[width=0.73\linewidth]{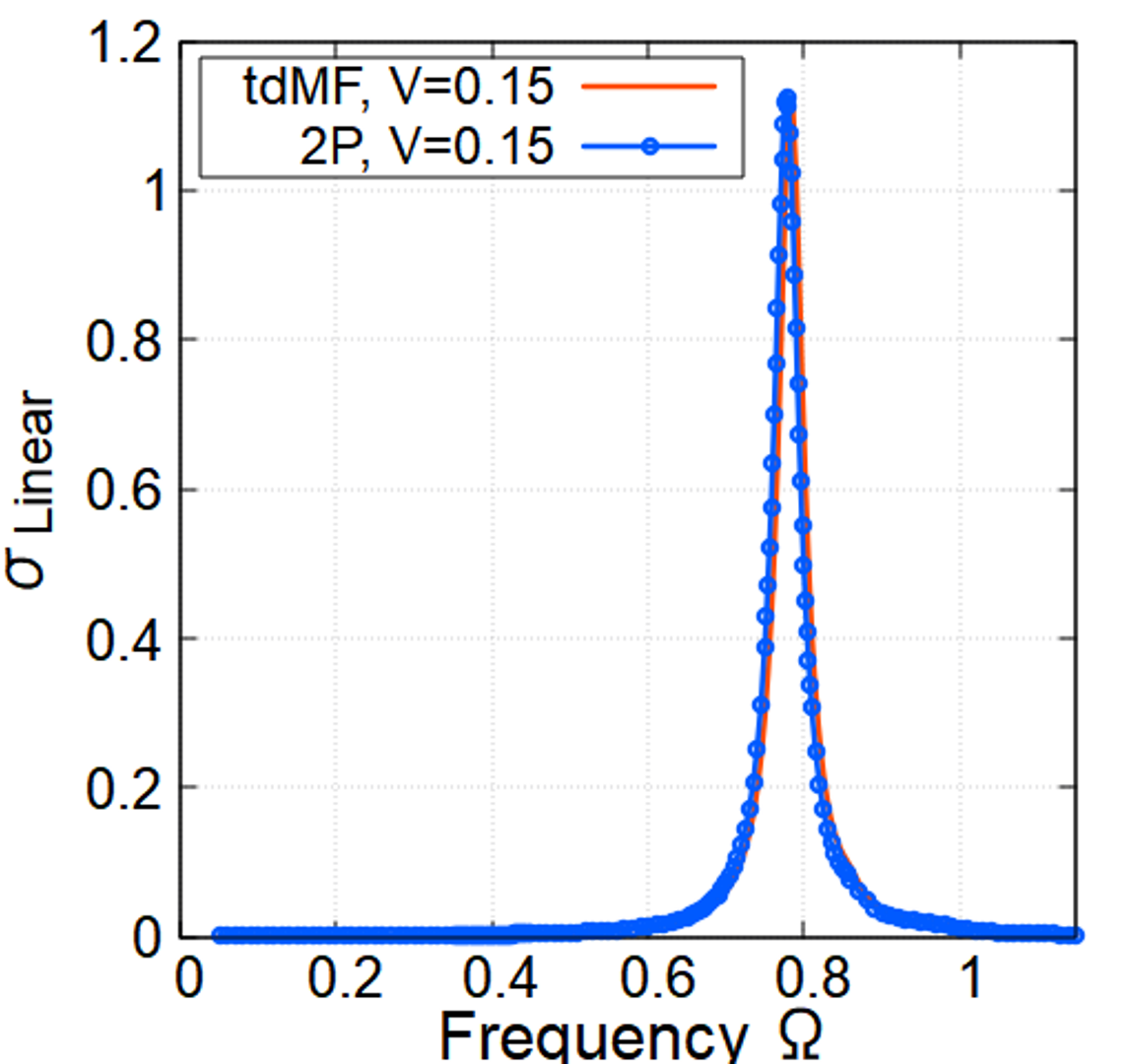}
\end{center}
\caption{Linear conductivity calculated by the correlation expansion including two-particle correlations (2P) and by tdMF for $V=0.03$ (upper panel) and $V=0.15$ (lower panel).}
\label{linear_cond_2P_effect}
\end{figure}

In Fig.~\ref{linear_cond_2P_effect}, we compare the linear conductivity for $V=0.03$ and $0.15$ calculated by tdMF and the correlation expansion, including two-particle correlation effects. The upper panel of Fig.~\ref{linear_cond_2P_effect} shows the results for weak interaction, $V=0.03$, and the lower panel shows $V=0.15$. For $V=0.03$, both approximations yield almost identical results. Two-particle correlation effects are not visible. On the other hand, for  $V=0.15$, two-particle correlations slightly affect the linear conductivity. While the height of the peak does not change, the spectrum is slightly shifted toward lower frequencies when including two-particle correlation effects. 
However, even for $V=0.15$, the impact of two-particle correlations on the linear response is small. We thus see that while interactions affect the linear response in this model, a time-dependent mean-field description of the system is sufficient to analyze the linear conductivity.

\begin{figure}[t]
\begin{center}
\includegraphics[width=0.73\linewidth]{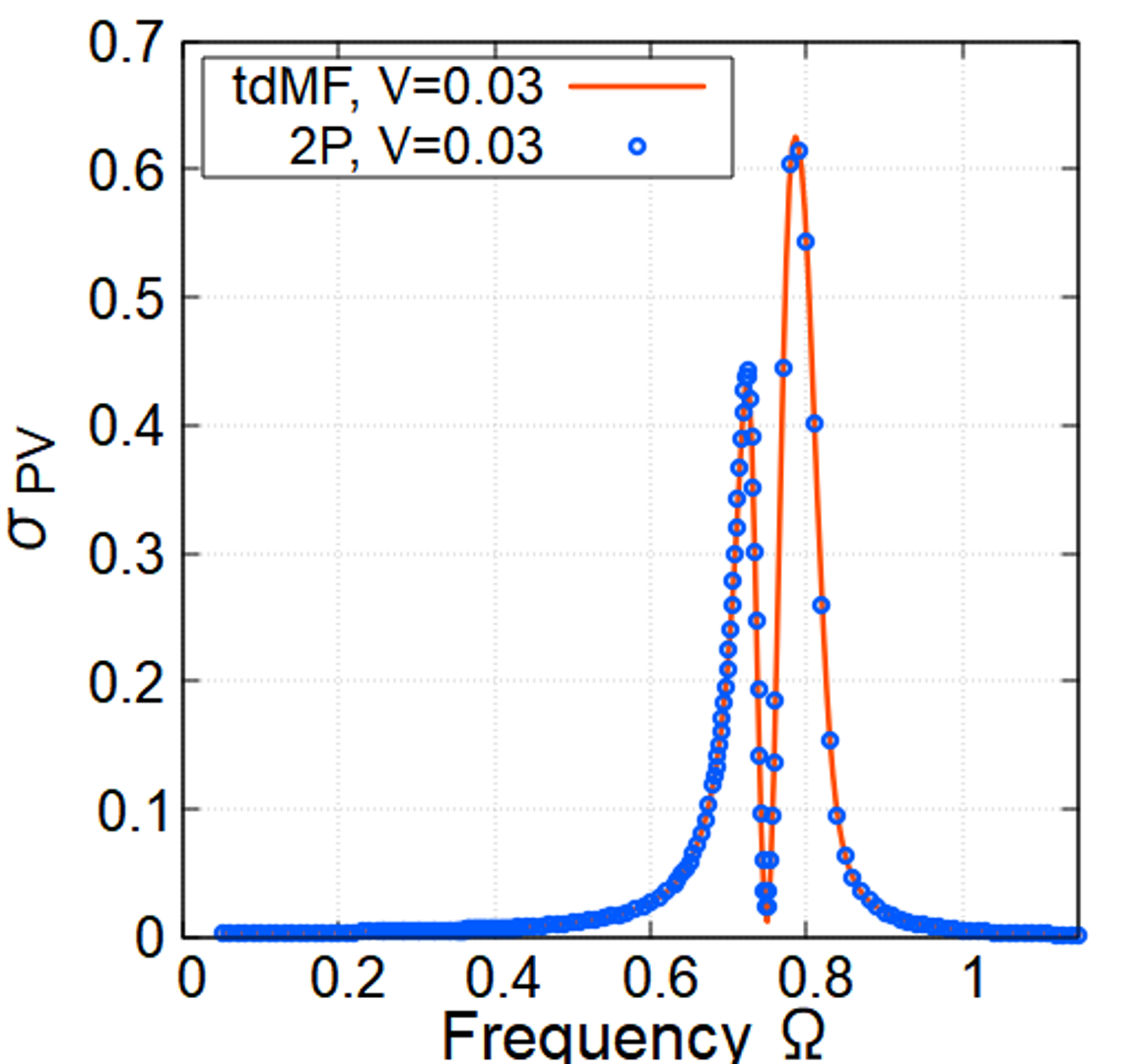}
\\
\includegraphics[width=0.73\linewidth]{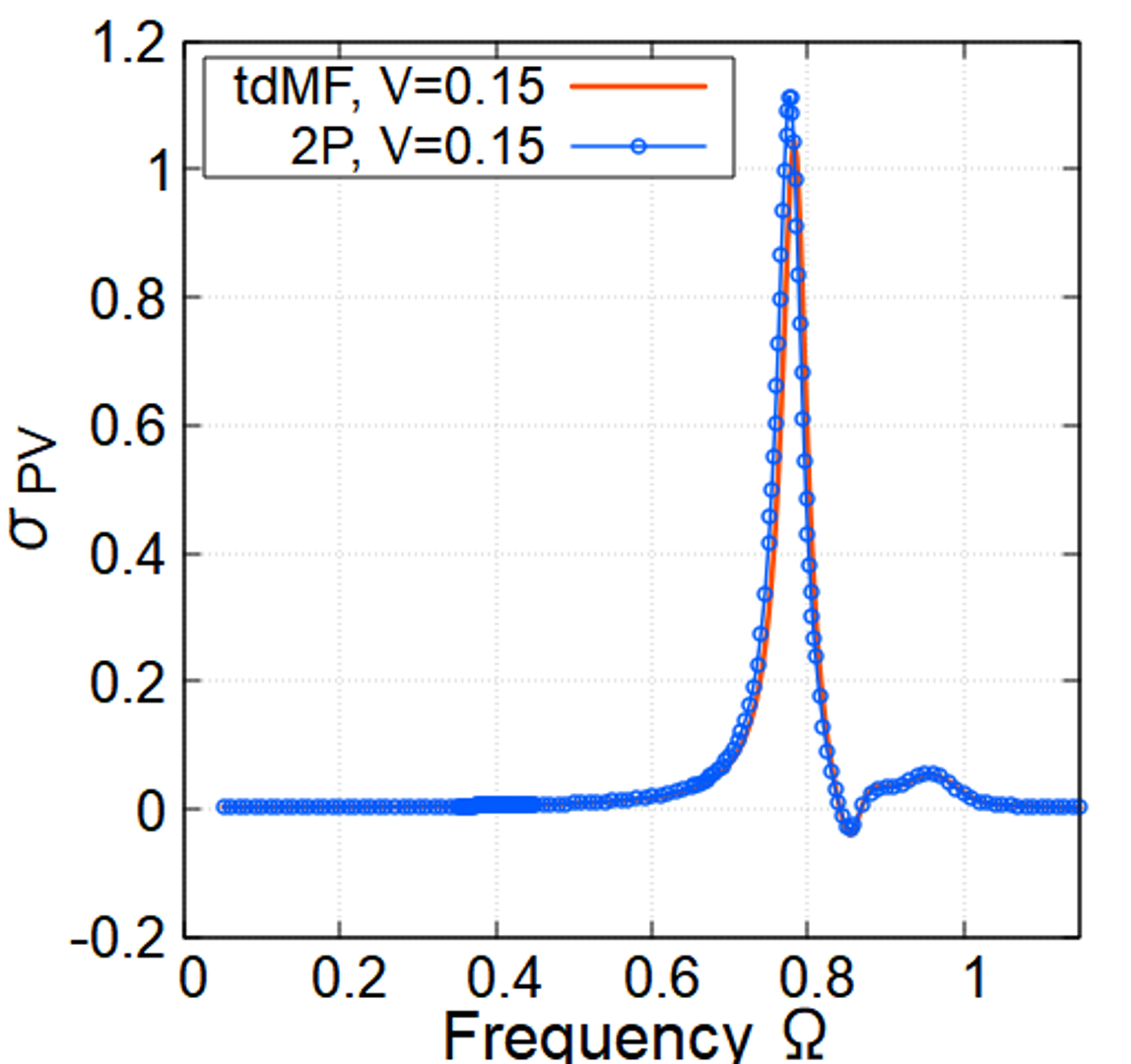}
\end{center}
\caption{Photovoltaic conductivity calculated by the correlation expansion (2P) and by tdMF for $V=0.03$ (upper panel) and $V=0.15$ (lower panel).}
\label{shift_cond_2P_effect}
\end{figure}

Next, we analyze two-particle correlation effects on the photovoltaic conductivity. Figure~\ref{shift_cond_2P_effect} shows the results for the photovoltaic conductivity calculated by tdMF and the correlation expansion for $V=0.03$ and $V=0.15$.
We note that the photovoltaic conductivity over the frequency mainly consists of two peaks: one peak around $\Omega = 0.7\sim 0.8$ and another peak at larger frequencies.
The left peak at $\Omega = 0.7 \sim 0.8$ corresponds to the excitonic peak, which has been mentioned above. The peak at slightly larger frequencies corresponds to the contribution from electrons that do not form excitons, which is confirmed in Appendix~\ref{appendix_electron_hole_correlation}.
A comparison between $V=0.03$ and $V=0.15$ immediately shows that interactions have a strong impact on the photovoltaic conductivity. The excitonic peak at $\Omega = 0.7 \sim 0.8$ is strongly enhanced from $\sigma_{\mathrm{PV}}\approx 0.45$ at $V=0.03$ to $\sigma_{\mathrm{PV}}\approx 1.12$ at $V=0.15$. On the other hand, the peak at higher frequencies is strongly suppressed.
Comparing tdMF and the correlation expansion up to second-order, we see that both spectra agree well. 
Two-particle correlations slightly enhance the magnitude of the excitonic peak. Furthermore, two-particle correlations slightly shift the spectrum for $V=0.15$. Compared to the linear conductivity, we can say that interactions have a strong effect on the photovoltaic conductivity. However, the impact of two-particle correlations beyond the mean-field level on the photovoltaic conductivity, although slightly stronger than on the linear conductivity, remains small.

\begin{figure}[t]
\begin{center}
\includegraphics[width=0.73\linewidth]{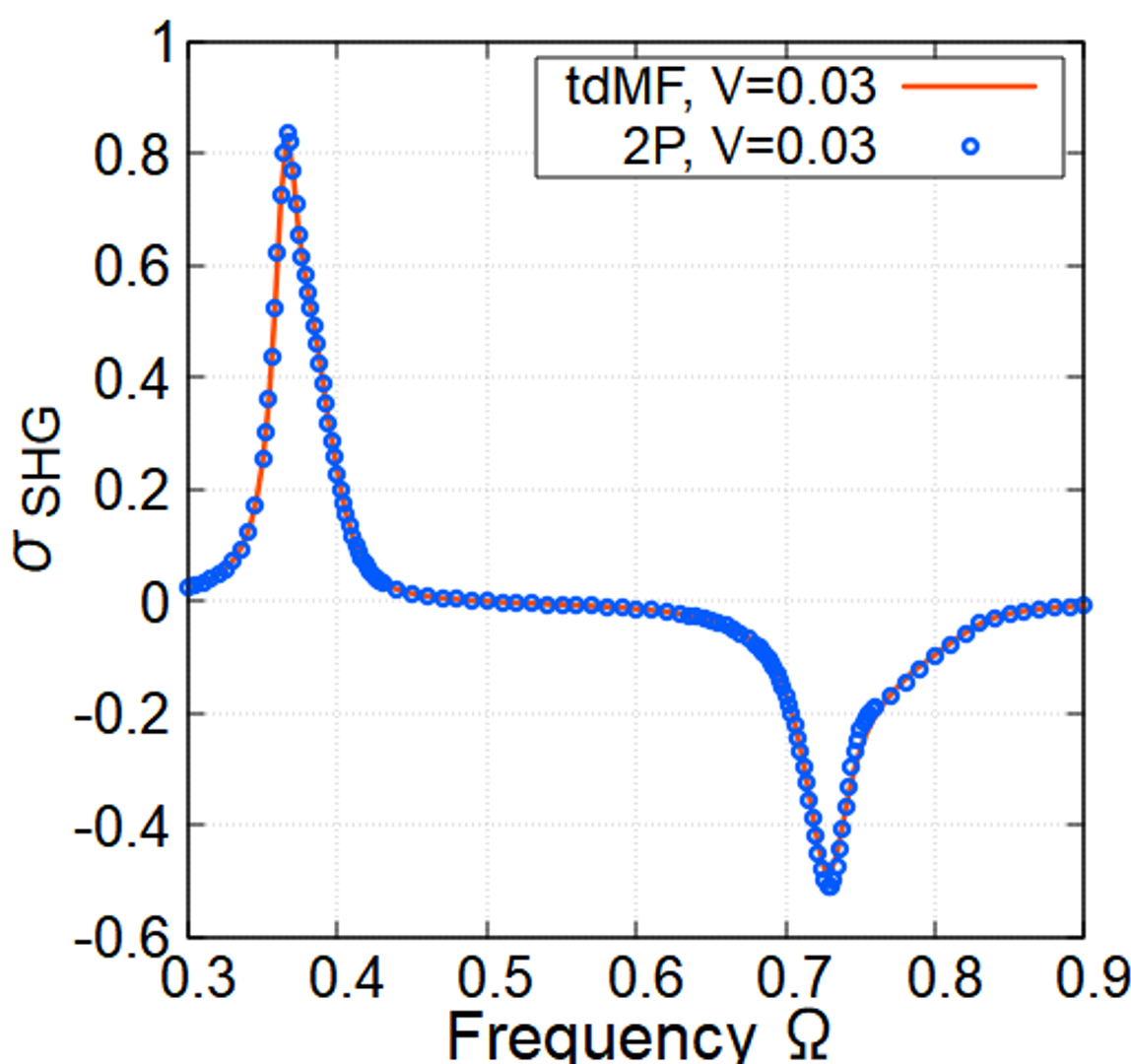}
\\
\includegraphics[width=0.73\linewidth]{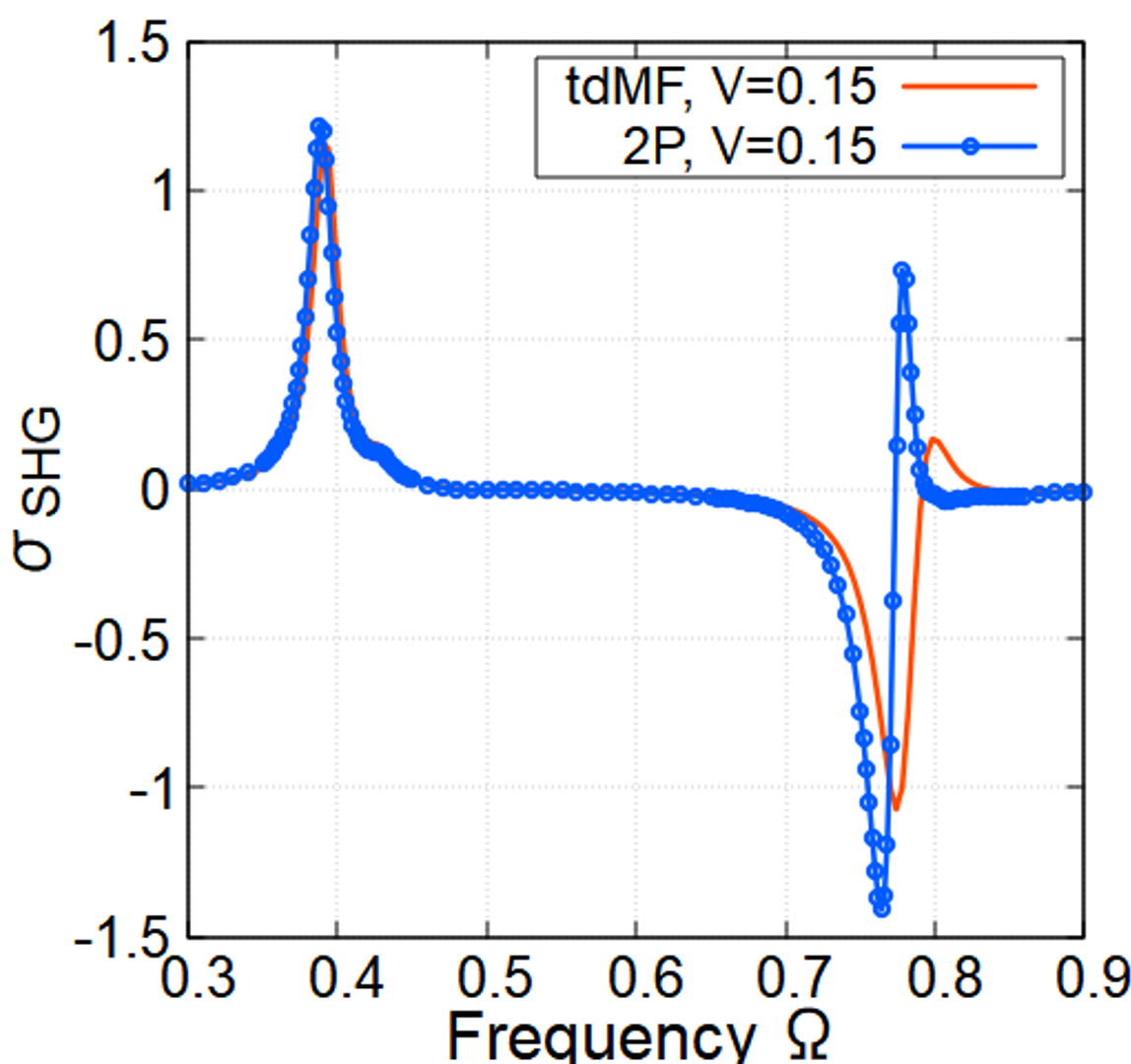}
\end{center}
\caption{SHG conductivity calculated by the correlation expansion (2P) and tdMF  for $V=0.03$ (upper panel) and $V=0.15$ (lower panel).}
\label{shg_cond_2P_effect}
\end{figure}

Then, we analyze the two-particle correlation effects on the SHG conductivity. In Fig.~\ref{shg_cond_2P_effect}, we show the SHG conductivity for $V=0.03$ and $V=0.15$ with and without two-particle correlation effects. Both figures demonstrate that the spectrum of the SHG conductivity consists of two peaks. The peak at low frequencies corresponds to two-photon excitations, which do not exist in the other spectra shown above.
The peak at high frequencies is a one-photon peak. Remarkably, this one-photon peak shows a strong dependence on the interaction strength.
For $V=0.03$ (upper panel of Fig.~\ref{shg_cond_2P_effect}), tdMF can describe the SHG conductivity well, and two-particle correlation effects are not very important. On the other hand, for $V=0.15$, the one-photon peak is significantly affected by two-particle correlations beyond the mean-field level. The spectrum is shifted toward low frequencies. Furthermore, the response is clearly enhanced by two-particle correlations, as can be seen at $\Omega\approx 0.8$ comparing between the red and blue lines.
In particular, the small positive peak becomes much sharper, and the height is about five times larger when two-particle correlations are included. We note that the two-photon peak at $\Omega\approx 0.4$ is not strongly affected by two-particle correlations. 
This is partly because excitations of electron-hole pairs around this peak originate from two-photon processes, including virtual excitations to intermediate states, so fewer electron-hole pairs are excited than around the one-photon peak.

\begin{figure}[t]
\begin{center}
\includegraphics[width=0.73\linewidth]{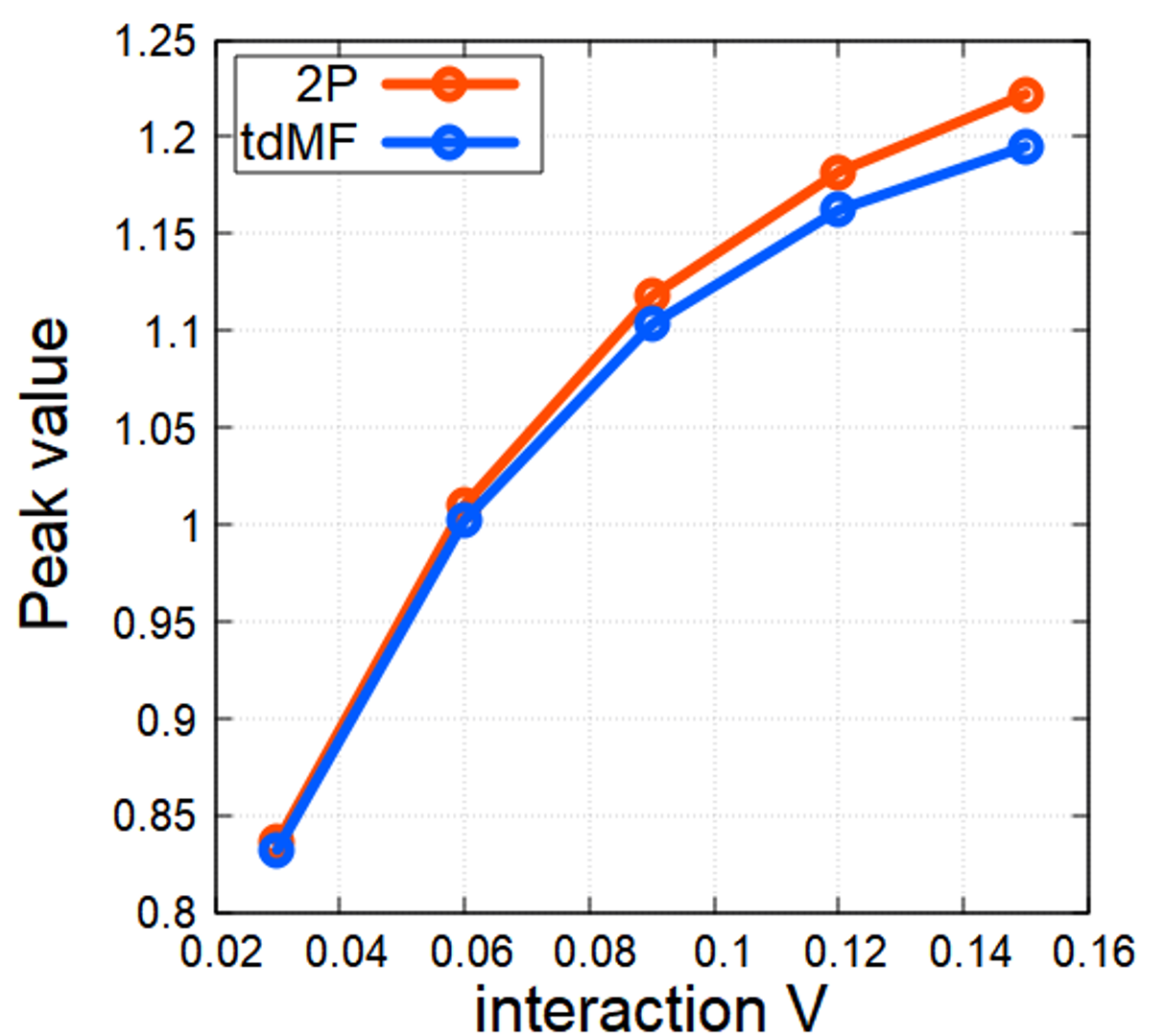}
\\
\includegraphics[width=0.73\linewidth]{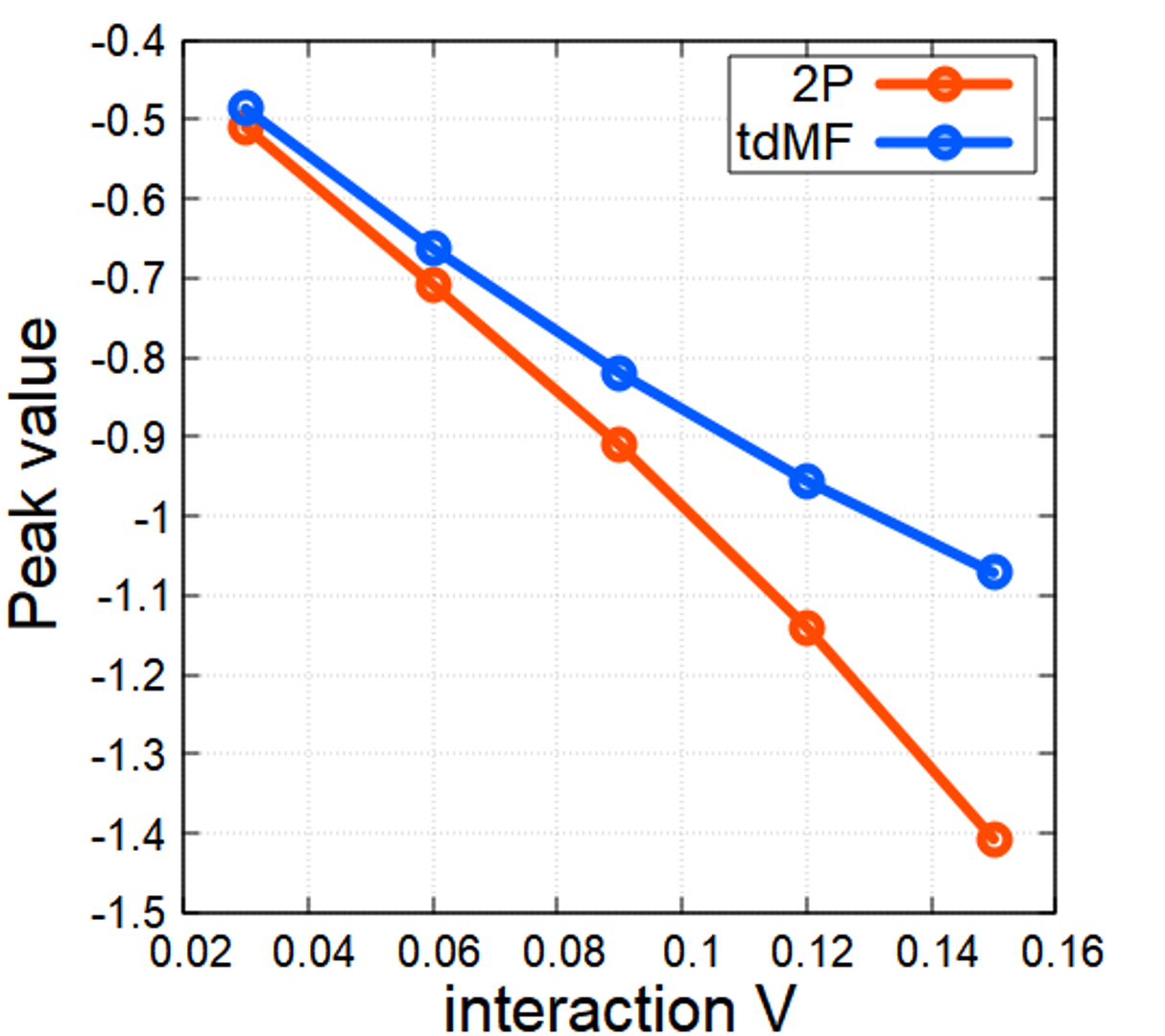}
\end{center}
\caption{Interaction strength dependence of the peak intensity in the SHG conductivity calculated by the correlation expansion (2P) and by tdMF for the two-photon peak (upper panel) and the one-photon peak(lower panel). The peak intensity of the one-photon peak corresponds to the negative peak in Fig.~\ref{shg_cond_2P_effect}.}
\label{shg_peak_V_dep}
\end{figure}

Finally, to analyze the impact of interactions on the SHG conductivity, in Fig.~\ref{shg_peak_V_dep}, we show the maximum value of the two-photon peak (left peak in Fig.~\ref{shg_cond_2P_effect}) in the upper panel and the maximum value of the one-photon peak (right peak in Fig.~\ref{shg_cond_2P_effect}) in the lower panel for different interaction strengths.
 We note that the maximum value of the one-photon peak is taken from the large negative peak, i.e., $\Omega \approx 0.72$ in the upper panel of Fig.~\ref{shg_cond_2P_effect}. 
 These results demonstrate that the SHG is strongly enhanced by the interaction, even on the mean-field level (tdMF). The strength of the SHG in the one-photon peak reaches a maximum value at $V=0.15$, which is nearly three times the maximum value of the non-interacting system.
 Also, as shown in the upper panel of Fig.~\ref{shg_peak_V_dep}, the interaction dependence of the two-photon peak is well described by the tdMF, which has been already demonstrated in Fig.~\ref{shg_cond_2P_effect}. On the other hand, two-particle correlations have a clear impact on the one-photon peak, which is further enhanced when taking into account two-particle correlations beyond the mean-field level.

In this section, we have seen that the impact of interactions and two-particle correlations on the linear response is weak compared to nonlinear conductivities.  
In particular, the one-photon peak of the SHG conductivity is significantly enhanced by two-particle correlations, and the shape of the spectrum is altered. We will explore these two-particle correlation effects in more detail in the next section.

\subsection{Two-particle correlations in the SHG conductivity}
\label{decomposition_of_SHG}
To analyze which two-particle correlation is essential for the current, we now decompose the current into single-particle contributions and six two-particle correlation terms, $\langle C^{\dagger}C^{\dagger}CC \rangle$, $\langle C^{\dagger}C^{\dagger}CV \rangle$, $\langle C^{\dagger}C^{\dagger}VV \rangle$, $\langle C^{\dagger}V^{\dagger}CV \rangle$, $\langle C^{\dagger}V^{\dagger}VV \rangle$, and $\langle V^{\dagger}V^{\dagger}VV \rangle$, as we have explained in Sec.~\ref{method_decomposition}.

\begin{figure}[t]
\begin{center}
\includegraphics[width=0.73\linewidth]{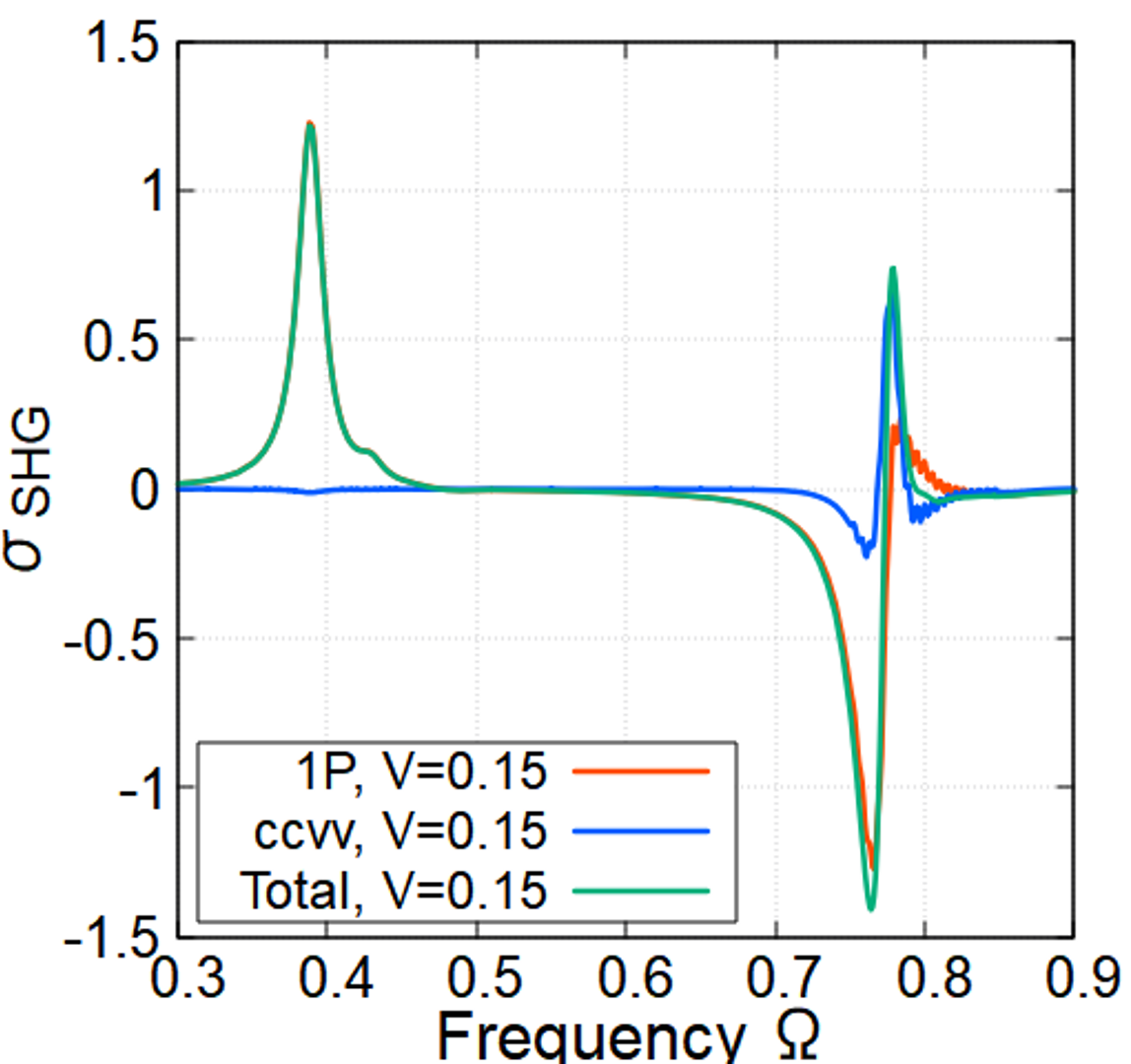}
\\
\includegraphics[width=0.73\linewidth]{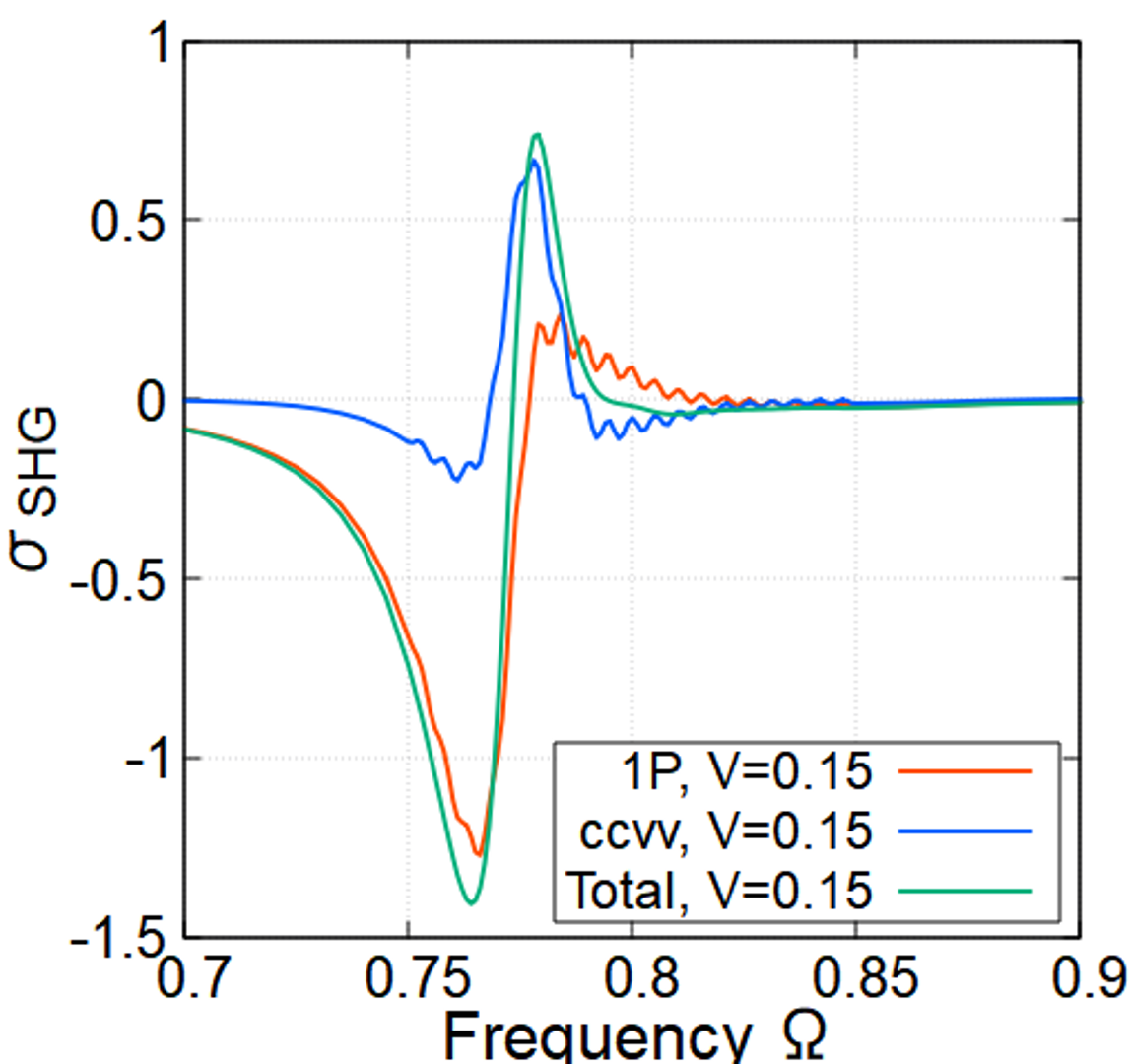}
\end{center}
\caption{Comparison between the one-particle contribution and $S_{ccvv}$ contribution to the SHG conductivity for $V=0.15$. The lower panel shows a magnification of the upper panel around $\Omega\sim0.7-0.9$.}
\label{shg_decomposition}
\end{figure}

Using this decomposition, we can see that the one-particle contribution and the two-particle term related to $\langle C^{\dagger}C^{\dagger}VV \rangle$  and $\langle V^{\dagger}V^{\dagger}CC \rangle$ (which are related by complex conjugation) are the dominant contributions to the SHG conductivity for $V=0.15$. In the upper panel of Fig.~\ref{shg_decomposition}, we show the total SHG conductivity, the one-particle, and $\langle C^{\dagger}C^{\dagger}VV \rangle$ contributions.
Other two-particle contributions are not shown here because they are small and almost negligible for this interaction strength.
The lower panel of Fig.~\ref{shg_decomposition} shows a magnification of the upper panel around the one-photon peak. 
As shown in the upper panel of Fig.~\ref{shg_decomposition}, two-particle contributions are tiny around the two-photon peak. Thus, one-particle contributions are almost identical to the total spectrum. This is consistent with the result in the previous section, showing that two-particle correlation effects on this peak are weak. Interactions, nevertheless, are important and enhance the response at these frequencies, as shown in the upper panel of Fig.~\ref{shg_peak_V_dep}. However, these results show that a  mean-field description is sufficient to analyze the two-photon peak at weak to moderate interaction strengths.
On the other hand, the contribution of $\langle C^{\dagger}C^{\dagger}VV \rangle$ constitutes a large fraction of the full spectrum for the one-photon peak. Notably, the sharp peak around $\Omega\sim0.78$ mainly originates from the $\langle C^{\dagger}C^{\dagger}VV \rangle$ contribution, which cannot be captured by only considering one-particle contributions and taking interactions into account only on the mean-field level.

We note that $\langle C^{\dagger}C^{\dagger}VV \rangle$ is similar to $\langle C^{\dagger}V \rangle$ in that it becomes finite when the system includes electron-hole pairs.
 If we consider $\langle c_{k+q,c}^{\dagger}c_{k^{'}-q,c}^{\dagger}c_{k^{'},v}c_{k,v} \rangle$ on the mean-field level, each electron-hole pair must have the same momentum, $k$ and $-k$, because of momentum conservation. On the other hand, $S_{ccvv}(k,k^{'},q)$ is the deviation from this mean-field expectation value. It can be finite even when electron-hole pairs have different momenta $k+q$ and $-k$. Thus, the number of possibilities to form excited electron-hole pairs is increased when including two-particle correlations, which results in an enhancement of the excitonic peak.

\subsection{Nonlinearity of $S_{ccvv}(k,k^{'},q)$}
\label{results_nonlinearity}
\begin{figure}[t]
\begin{center}
\includegraphics[width=0.73\linewidth]{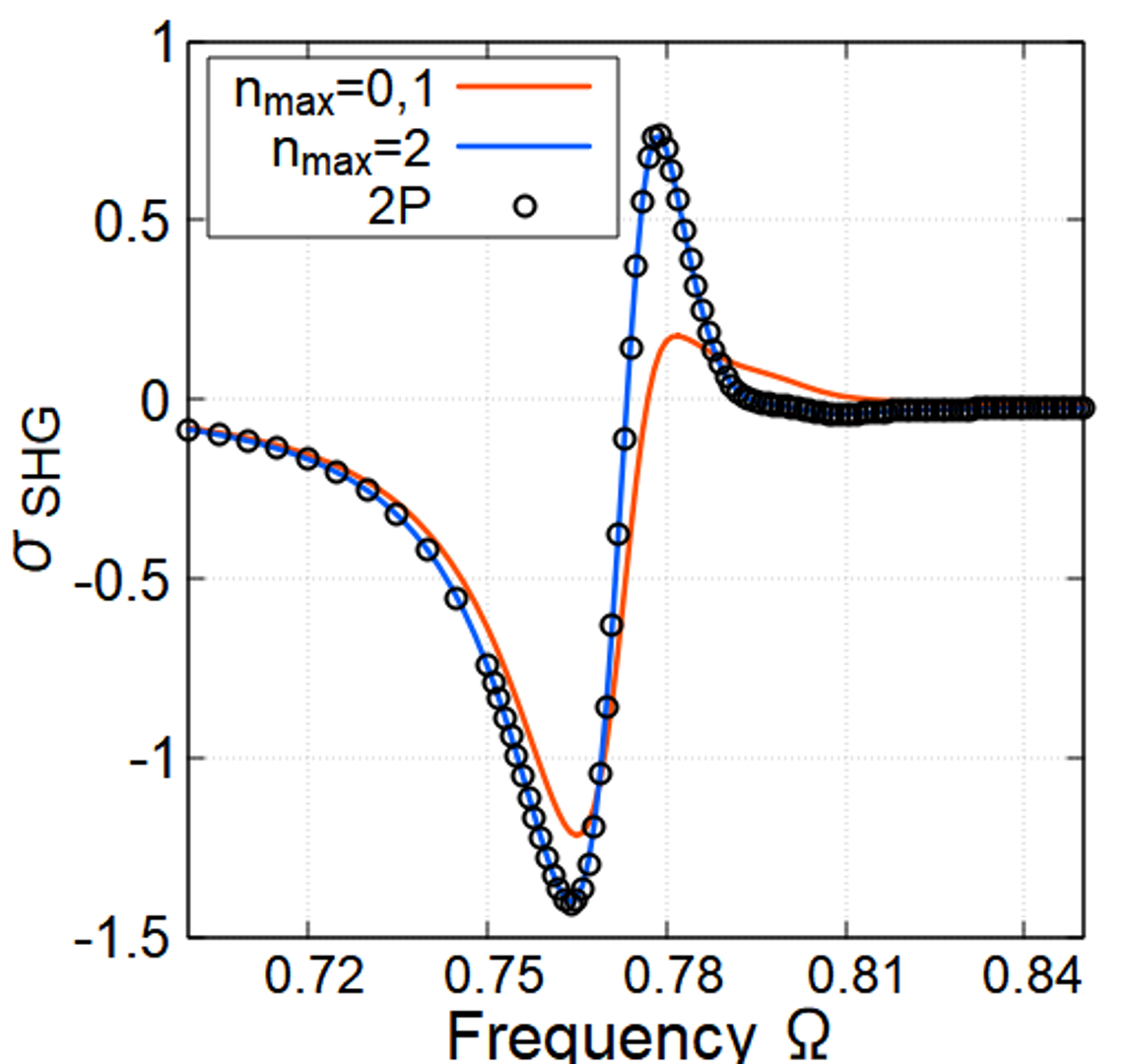}
\end{center}
\caption{Contributions of different harmonics ($n_{\mathrm{max}}=0,1,2,\infty$) to the one-photon peak in the SHG conductivity for $V=0.15$.}
\label{ccvv_approximation}
\end{figure}

In Sec.~\ref{decomposition_of_SHG}, we have revealed that the contribution related to $\langle C^{\dagger}C^{\dagger}VV \rangle$ is essential in understanding the SHG conductivity. In this section, we analyze the dynamics of $S_{ccvv}(k,k^{'},q)$. 
Furthermore, we study whether the time-dependence of  $S_{ccvv}(k,k^{'},q)$ is important to understand the SHG conductivity.
 To analyze this point,  we first Fourier transform $S_{ccvv}(k,k^{'},q)$, using the non-equilibrium steady state:
\begin{equation}
\label{Sccvv_fourier}
S_{ccvv}(k,k^{'},q)=\sum_{n=-\infty}^{\infty}\tilde{S}_{ccvv}^{n\Omega}(k,k^{'},q)e^{-in\Omega t}.
\end{equation}
We then consider approximations of $S_{ccvv}(k,k^{'},q)$, by truncating the summation at $n_{\mathrm{max}}$ as:
\begin{equation}
\begin{aligned}
S_{ccvv}^{n_{\mathrm{max}}}(k,k^{'},q) = \sum_{n=-n_{\mathrm{max}}}^{n_{\mathrm{max}}}\tilde{S}_{ccvv}^{n\Omega}(k,k^{'},q)e^{-in\Omega t}.
\end{aligned}
\end{equation}
 For example, if we truncate at $n_{\mathrm{max}}=2$, we ignore 3rd and higher-order harmonics in this two-particle correlation function.
 Using the approximated $S_{ccvv}^{n_{\mathrm{max}}}(k,k^{'},q)$ (and the same approximation for $S_{vvcc}^{n_{\mathrm{max}}}(k,k^{'},q)$), we can recalculate the current and the SHG conductivity varying $n_{\mathrm{max}}$ in Eq.~(\ref{decomposed_current_int}). 
 In Fig.~\ref{ccvv_approximation}, we show the SHG conductivity for $n_{\mathrm{max}}=0,1,2$ and the full conductivity($n_{\mathrm{max}}=\infty$) for $V=0.15$. We note that the calculation for $n_{\mathrm{max}}=0$ and $n_{\mathrm{max}}=1$ yield identical results; thus, they are shown together in this figure. From this result, we see that there is no $1\Omega$ contribution from $S_{ccvv}(k,k^{'},q)$ and $S_{vvcc}(k,k^{'},q)$ to the SHG conductivity. 
 Furthermore, as can be seen in this figure, the conductivity for $n_{\mathrm{max}}=2$ completely reproduces the full conductivity. This is very natural because third-order harmonics correspond to at least third-order perturbations in the electric field. Thus, when calculating the SHG conductivity, $S_{ccvv}^{2}(k,k^{'},q)$ and $S_{vvcc}^{2}(k,k^{'},q)$ are sufficient. 
On the other hand, the conductivity for $n_{\mathrm{max}}=1$ deviates from the full conductivity. This difference is especially large around $\Omega \sim 0.78$, where two-particle correlation effects are important as shown in Fig.~\ref{shg_cond_2P_effect} and \ref{shg_decomposition}. Thus, Fig.~\ref{ccvv_approximation} reveals that two-particle correlation effects in our calculations are related to second-order harmonics of $S_{ccvv}(k,k^{'},q)$ and $S_{vvcc}(k,k^{'},q)$ with frequency $2\Omega$.

\subsection{Enhancement of two-particle correlations}
\label{results_enhancement_correlations}
\begin{figure}[t]
\begin{center}
\includegraphics[width=0.73\linewidth]{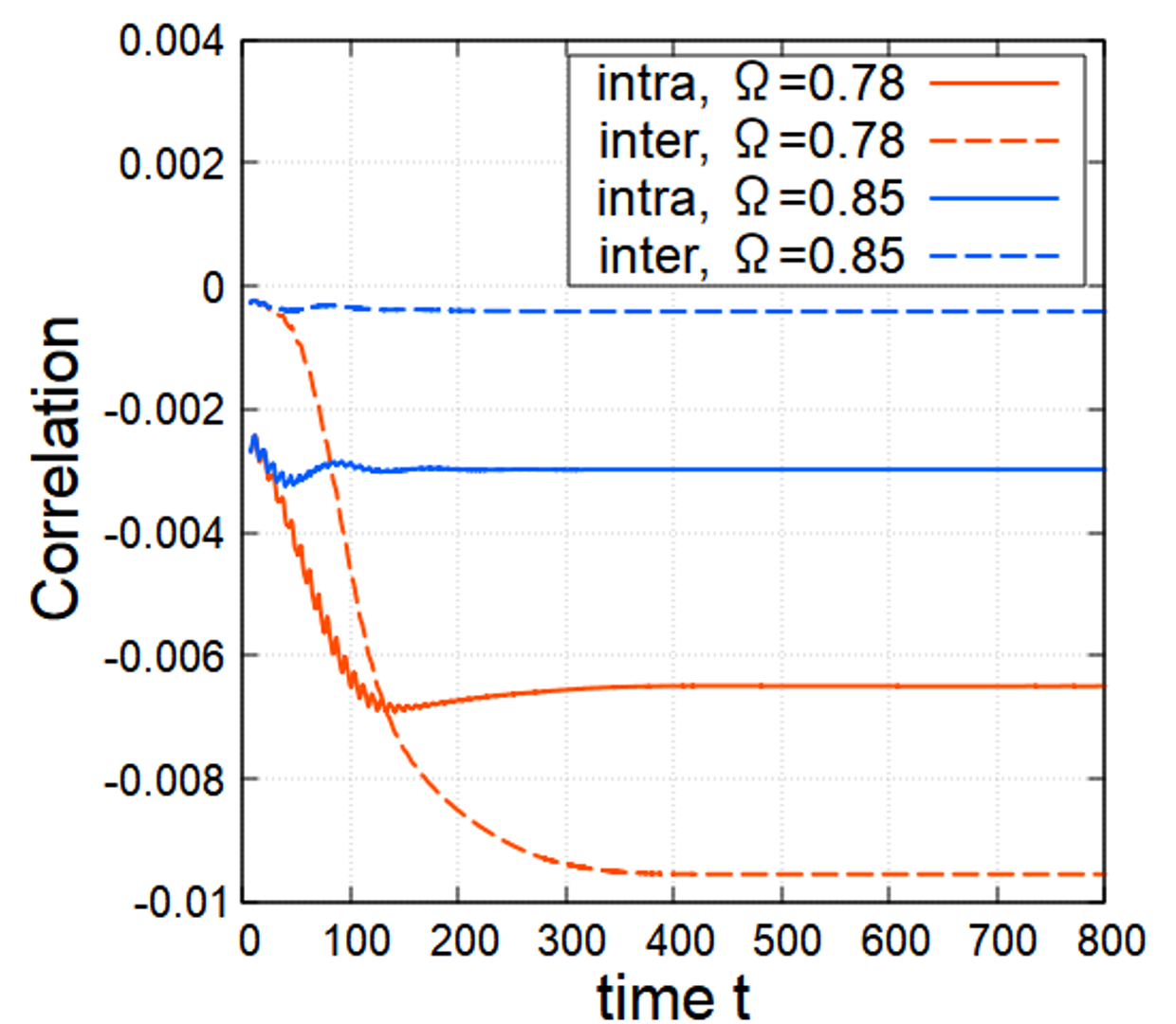}
\\
\includegraphics[width=0.73\linewidth]{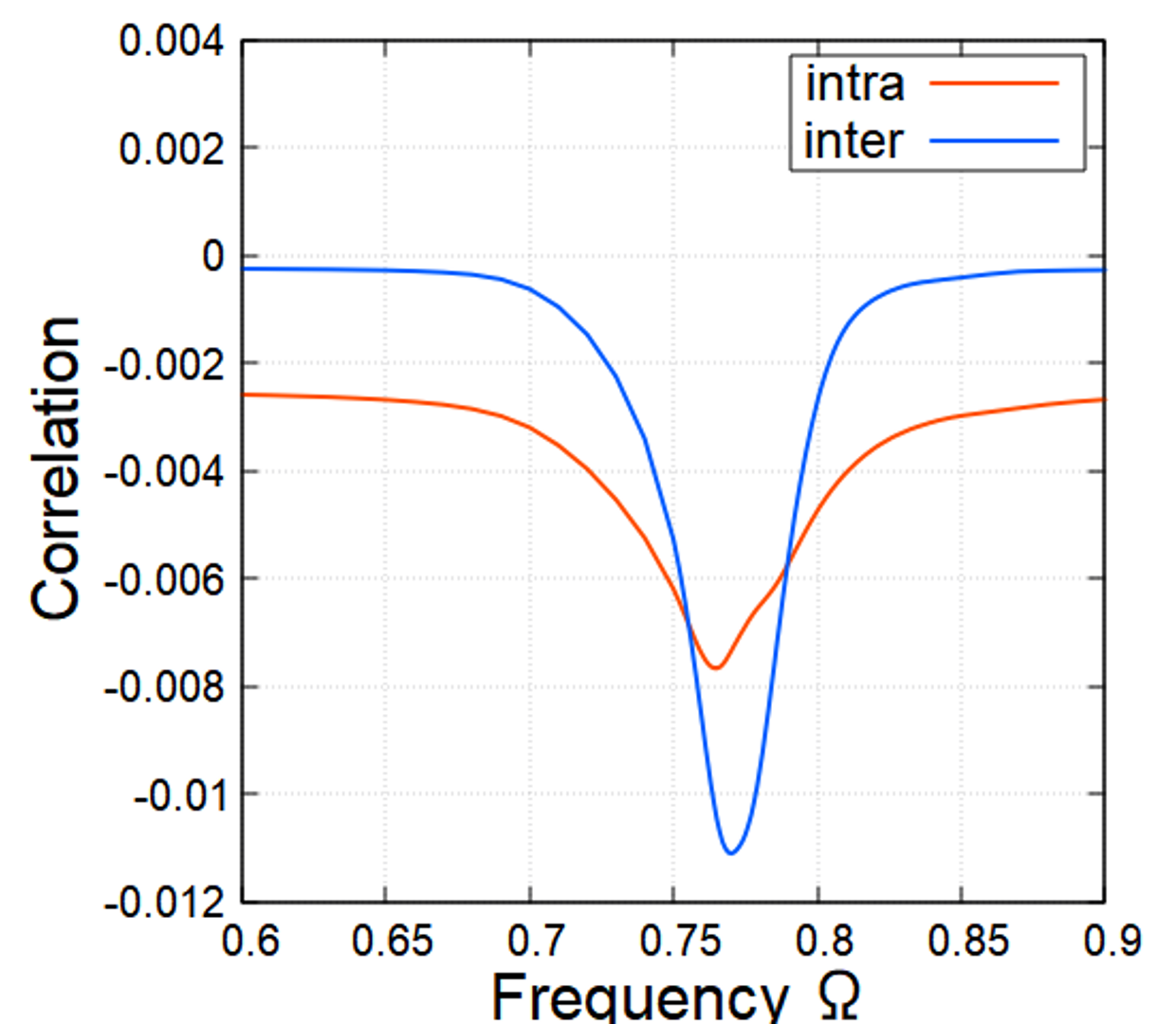}
\end{center}
\caption{(upper panel): Time dependence of the averaged intercell and intracell density-density correlation for $\Omega=0.78$ and $\Omega=0.85$. (lower panel): Comparison of intercell and intracell correlations in the steady state for different frequencies. Around the excitonic peak, the intercell correlations are strongly enhanced and exceed the intracell correlations. }
\label{intra_inter_correlation}
\end{figure}

In this section, we study the charge-charge correlations in the non-equilibrium steady state. We use an electric field with amplitude $E_{0}=-0.05$ in this section, assuming a strongly driven correlated electron system. The interaction is set to $V=0.15$. In the upper panel of Fig.~\ref{intra_inter_correlation}, we show the time evolution of the intracell charge-charge correlations $\langle n_{i,A}n_{i,B} \rangle^{c}$ and the intercell charge-charge correlations $\langle n_{i+1,A}n_{i,B} \rangle^{c}$ under the external electric field for $\Omega=0.78,0.85$. The frequency $\Omega=0.78$ corresponds to the excitonic peak as shown in Fig.~\ref{linear_cond_2P_effect}. $\Omega=0.85$ is slightly above the excitonic peak. We note that the plotted values are averaged over one period,
\begin{equation}
\begin{aligned}
\langle n_{i,A}n_{i,B} \rangle_{av}^{c} &= \frac{\Omega}{2\pi}\int_{t-2\pi/\Omega}^{t}dt^{'} \langle n_{i,A}n_{i,B} \rangle^{c} \\
\langle n_{i+1,A}n_{i,B} \rangle_{av}^{c} &= \frac{\Omega}{2\pi}\int_{t-2\pi/\Omega}^{t}dt^{'} \langle n_{i+1,A}n_{i,B} \rangle^{c}
\end{aligned}
\end{equation}
As seen in the upper panel of Fig.~\ref{intra_inter_correlation}, the change of charge-charge correlations due to the electric field switched on at $t=0$ is small at $\Omega=0.85$. On the other hand, charge-charge correlations are significantly enhanced for $\Omega=0.78$. The enhancement of the intercell correlations for $\Omega=0.78$ is so large that the absolute value of the intercell correlation exceeds the intracell correlations. 
This enhancement is further analyzed in the lower panel of Fig.~\ref{intra_inter_correlation}, which shows the frequency dependence of the enhancement in the steady state. We see that both correlations are enhanced only around the excitonic peak. The intercell correlations exceed the intracell correlations at the excitonic peak. These results
demonstrate that correlations and fluctuations in the non-equilibrium state can be qualitatively different from those in equilibrium.

\section{Conclusion}
In summary, we have calculated linear and nonlinear responses in a 1d Rice-Mele model, including two-particle correlation effects. Our approach is based on the correlation expansion method, which enables us to calculate non-equilibrium states by simulating the time evolution of the one-particle and two-particle density matrices. We have analyzed the impact of interactions and, particularly, two-particle correlations on the linear conductivity, the photovoltaic effect, and the SHG conductivity.
We have shown that the conductivity at the frequency corresponding to the excitonic excitation in this system is enhanced by interactions.
However, while we have seen that interactions affect the linear conductivity, we have demonstrated that two-particle correlation effects beyond the mean-field level are more salient in nonlinear conductivities. 
Notably, the one-photon peak in the SHG conductivity is significantly enhanced by two-particle correlations beyond the mean-field level.
To understand which two-particle correlations affect the conductivities most strongly, we have decomposed the current into one-particle and two-particle contributions. Utilizing this decomposition, we have revealed that $S_{ccvv}(k,k^{'},q)$ and $S_{vvcc}(k,k^{'},q)$ are the most important contributions to the enhancement of the SHG conductivity. In addition, we have shown that the second-order harmonics of $S_{ccvv}(k,k^{'},q)$ and $S_{vvcc}(k,k^{'},q)$ are essential, which cannot be treated within methods only considering the dynamics of one-particle quantities. Finally, we have calculated the real-time dynamics of charge-charge correlations. We have seen that the two-particle correlations are enhanced by the external electric field with frequency around the excitonic peak. Furthermore, external driving significantly enhances the intercell two-particle correlations so that the magnitude of intercell correlations can exceed intracell two-particle correlations. 

Our calculations demonstrate that two-particle correlations can affect nonlinear conductivities considerably and cannot be ignored when assessing nonlinear responses.
Even considering only electron-hole systems, two-particle correlation effects include various many-body phenomena, such as the impact ionization, the Auger recombination, and other excitonic effects. Controlling these effects is known to be essential to realizing efficient solar cells\cite{manousakis_2010,CHANTANA2021342}. Our approach enables us to calculate nonlinear optical properties, including two-particle correlation effects at a microscopic level. Furthermore, our method is a real-time approach and easily extended to simulate optical responses under a pump pulse setup. Such calculations would be future problems, which lead to a deeper understanding of photoexcited correlated electron systems.
\label{conclusion}

\section*{Acknowledgements}
\label{acknowledgements}
We thank Kento Uchida and Koki Shinada for their insightful discussions.
R.P. is supported by JSPS KAKENHI No.~JP23K03300.  This work was supported by JST, the establishment of university fellowships towards the creation of science and technology innovation, and Grant Number JPMJFS2123.
Parts of the numerical simulations in this work have been done using the facilities of the Supercomputer Center at the
Institute for Solid State Physics, the University of Tokyo.

\appendix
\section{Interacting terms in the tdMF}
\label{M_definition}
Here, we show the explicit expression for $M_{\alpha\beta}$ in Eq.~(\ref{MF_EOM}) of the tdMF equations. They are defined as:
\begin{equation}
\begin{aligned}
M_{cc}(k)&=\sum_{k^{'}}F_{cccc}(k_{t}^{'},k_{t},0)f_{c}(k^{'})\\
&+\sum_{k^{'}}F_{cccc}(k_{t},k_{t}^{'},0)f_{c}(k^{'})\\
&-\sum_{k^{'}}F_{cccc}(k_{t},k_{t}^{'},k_{t}^{'}-k_{t})f_{c}(k^{'})\\
&-\sum_{k^{'}}F_{cccc}(k_{t}^{'},k_{t},k_{t}-k_{t}^{'})f_{c}(k^{'})\\
&+\sum_{k^{'}}F_{cccv}(k_{t}^{'},k_{t},0)y(k^{'})\\
&-\sum_{k^{'}}F_{cccv}(k_{t}^{'},k_{t},k_{t}-k_{t}^{'})y(k^{'})\\
&+\sum_{k^{'}}F_{cvcc}(k_{t},k_{t}^{'},0)y^{*}(k^{'})\\
&-\sum_{k^{'}}F_{cvcc}(k_{t}^{'},k_{t},k_{t}-k_{t}^{'})y^{*}(k^{'})\\
&-\sum_{k^{'}}F_{cvcv}(k_{t}^{'},k_{t},k_{t}-k_{t}^{'})f_{v}(k^{'})\\
\end{aligned}
\end{equation}
\begin{equation}
\begin{aligned}
M_{vv}(k)&=-\sum_{k^{'}}F_{cvcv}(k_{t},k_{t}^{'},k_{t}^{'}-k_{t})f_{c}(k^{'})\\
&+\sum_{k^{'}}F_{cvvv}(k_{t}^{'},k_{t},0)y(k^{'})\\
&-\sum_{k^{'}}F_{cvvv}(k_{t},k_{t}^{'},k_{t}^{'}-k_{t})y(k^{'})\\
&+\sum_{k^{'}}F_{vvcv}(k_{t},k_{t}^{'},0)y^{*}(k^{'})\\
&-\sum_{k^{'}}F_{vvcv}(k_{t},k_{t}^{'},k_{t}^{'}-k_{t})y^{*}(k^{'})\\
&+\sum_{k^{'}}F_{vvvv}(k_{t}^{'},k_{t},0)f_{v}(k^{'})\\
&+\sum_{k^{'}}F_{vvvv}(k_{t},k_{t}^{'},0)f_{v}(k^{'})\\
&-\sum_{k^{'}}F_{vvvv}(k_{t},k_{t}^{'},k_{t}^{'}-k_{t})f_{v}(k^{'})\\
&-\sum_{k^{'}}F_{vvvv}(k_{t}^{'},k_{t},k_{t}-k_{t}^{'})f_{v}(k^{'})
\end{aligned}
\end{equation}
\begin{equation}
\begin{aligned}
M_{cv}(k)&=\sum_{k^{'}}F_{cccv}(k_{t},k_{t}^{'},0)f_{c}(k^{'})\\
&-\sum_{k^{'}}F_{cccv}(k_{t},k_{t}^{'},k_{t}^{'}-k_{t})f_{c}(k^{'})\\
&+\sum_{k^{'}}F_{ccvv}(k_{t}^{'},k_{t},0)y(k^{'})\\
&+\sum_{k^{'}}F_{ccvv}(k_{t},k_{t}^{'},0)y(k^{'})\\
&-\sum_{k^{'}}F_{ccvv}(k_{t},k_{t}^{'},k_{t}^{'}-k_{t})y(k^{'})\\
&-\sum_{k^{'}}F_{ccvv}(k_{t}^{'},k_{t},k_{t}-k_{t}^{'})y(k^{'})\\
&+\sum_{k^{'}}F_{cvcv}(k_{t},k_{t}^{'},0)y^{*}(k^{'})\\
&+\sum_{k^{'}}F_{cvvv}(k_{t},k_{t}^{'},0)f_{v}(k^{'})\\
&-\sum_{k^{'}}F_{cvvv}(k_{t}^{'},k_{t},k_{t}-k_{t}^{'})f_{v}(k^{'})\\
M_{vc}(k) &= M_{cv}^{*}(k)
\end{aligned}
\end{equation}

\section{Initial state in the correlation expansion method}
\label{appendix_ground_state}
\begin{figure*}[t]
\label{adiabatic_switch}
\begin{center}
\includegraphics[width=0.325\linewidth]{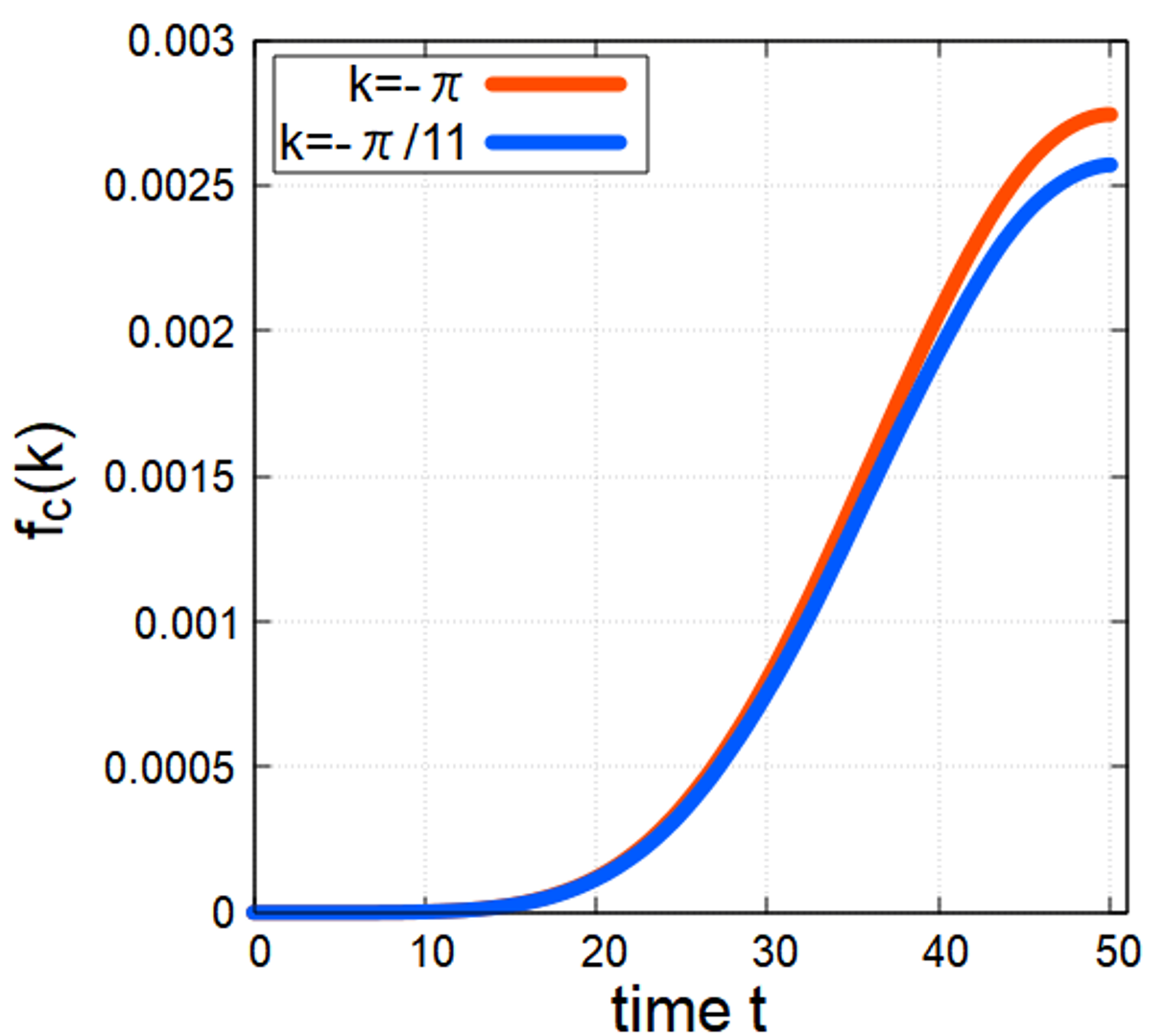}
\includegraphics[width=0.325\linewidth]{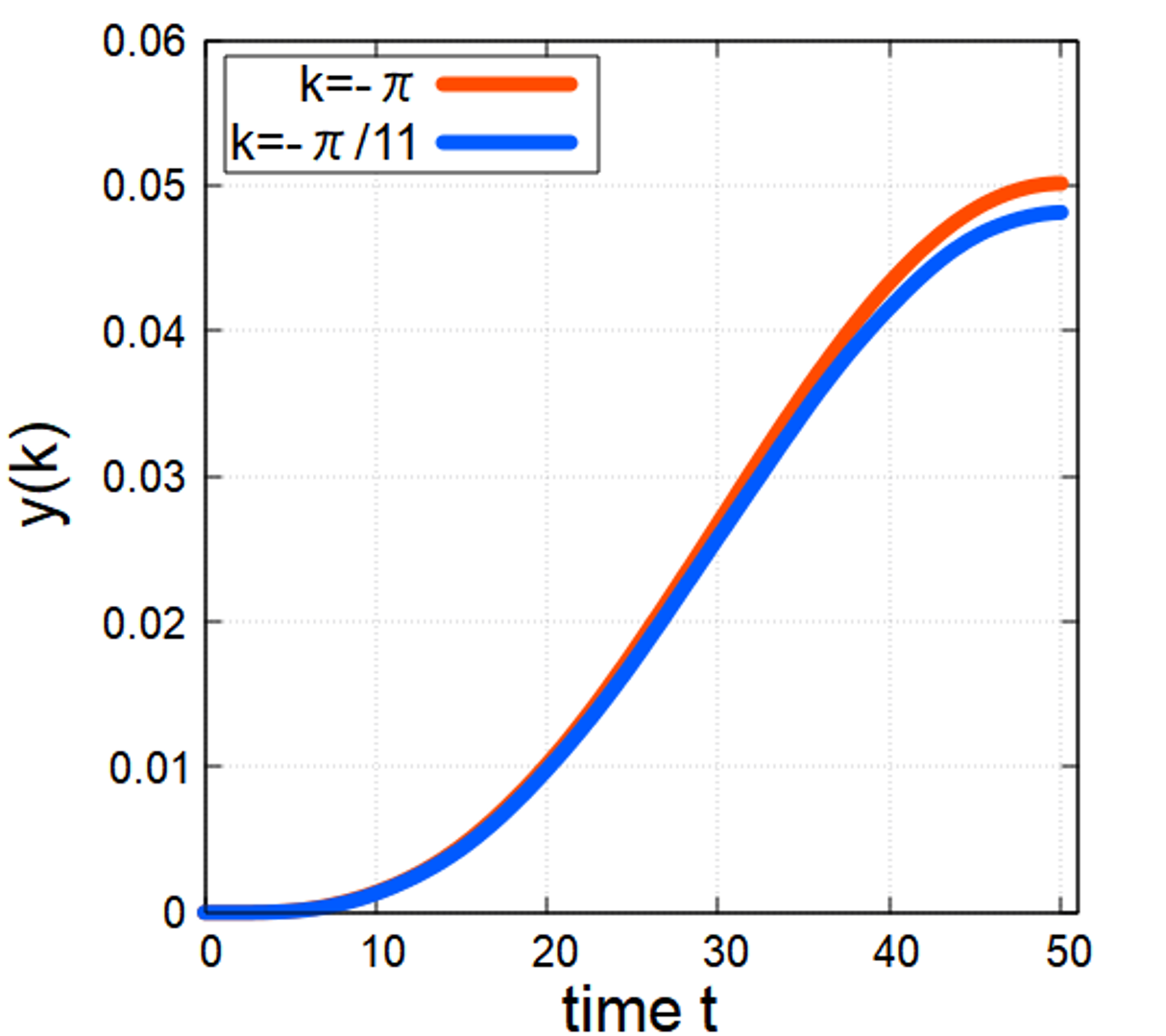}
\includegraphics[width=0.325\linewidth]{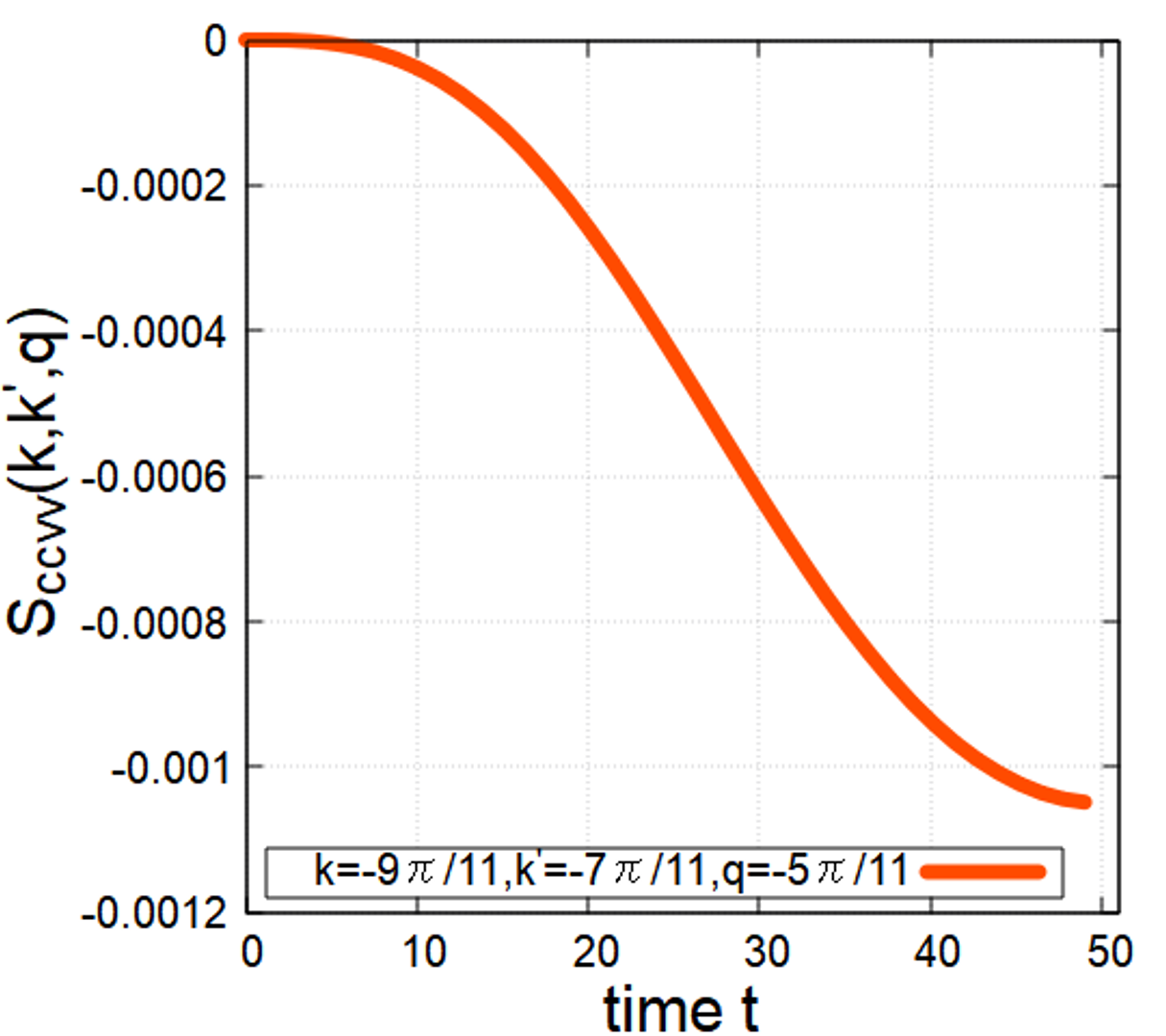}
\end{center}
\caption{One-particle expectation values $f_c(k)$, $y(k)$, and two-particle correlation $S_{ccvv}$ during the adiabatic switching on of the interaction strength for $V=0.15$, $T=50$, and different momenta. 
All correlations reach a plateau at $t=T(=50)$ without large oscillations.}
\label{exciton_correlation}
\end{figure*}

To obtain the ground state within the correlation expansion, we adiabatically switch on the interaction strength and calculate the time evolution of the density matrices. The interaction is given as
\begin{equation}
\label{time_dependence_of_interaction}
\begin{aligned}
V(t) = V \sin^{3} \left( \frac{\pi t}{2 T} \right),
\end{aligned}
\end{equation}
where $T$ determines the speed at which the interaction is switched on. In this paper, we use $T=50.0$. We use $V(t)$ in the EOM for the correlation expansion method without external electric field and relaxation, i.e., $E_{0}=0$ and $\gamma=0$.
We then calculate the time evolution starting from the non-interacting initial state at $t=0$ and adiabatically switch on the interaction.
We use the density matrices at $t=T$ as the approximate ground state and use them as the initial state in the calculations in the main text.
An example of this procedure for $V=0.15$ is shown in Fig.~\ref{adiabatic_switch}.
We see how the occupation in the conduction band, $f_c(k)$, and the transition element between conduction and valence band, $y(k)$,
 increase and reach stationary states at $t=T$ when switching on the interaction.
The interaction induces a change in the occupation numbers. While in the non-interacting system, only the valence band is occupied, in the interacting model, there are electrons in the conduction band. We also include the value of $S_{ccvv}$ in Fig.~\ref{adiabatic_switch}. This correlation function smoothly decreases to a negative value and reaches a plateau at $t=T$ when the interaction is turned on.
These results show that even at $V=0.15$, the occupation number in the conduction band, the transition element, and two-particle correlations behave smoothly without large oscillations and reach a plateau at $t=T$ when the interaction is turned on. Thus, we conclude that we can calculate an accurate ground state, which we use as the initial state in the real-time calculations.

\section{Electron-hole correlations in real space}
\label{appendix_electron_hole_correlation}
To demonstrate that our system has an excitonic nature and that the sharp peak of the spectrum in the main text corresponds to an excitonic peak, we calculate the electron-hole correlation function in real space under an external electric field. We consider the system excitonic if we can confirm that the conduction electrons and valence band holes are bound to each other at a small relative distance. We note that the calculations in this appendix are done by tdMF because excitonic properties appear even at the mean-field level, as shown in this section.

We define the creation(annihilation) operators for conduction electrons and valence electrons in real space $c_{i,c/v}^{\dagger}(c_{i,c/v})$ as the Fourier transform of those electrons in the momentum space $c_{k,c/v}^{\dagger}(c_{k,c/v})$. Using these operators, we calculate the correlation functions of the electron density in the conduction band at site $i$ and the hole density in the valence band at site $j$, $\langle n_{i,c}(1-n_{j,v}) \rangle$. We note that the basis used here differs slightly from that in the main text. While in the main text, the Houston basis is defined by the non-interacting part of the Hamiltonian, here, we absorb the interaction at the mean-field level into the Hamiltonian.
This change makes the interpretation of the results easier.
The one-particle terms in the Hamiltonian are  renormalized as follows:
\begin{equation}
\label{renormalization}
\begin{aligned}
Q_{x} &\rightarrow Q_{x} - V(\langle c_{i,B}^{\dagger} c_{i,A} \rangle + \langle c_{i+1,A}^{\dagger} c_{i,B} \rangle)\\
Q_{y} &\rightarrow Q_{x} + V(\langle c_{i,B}^{\dagger} c_{i,A} \rangle - \langle c_{i+1,A}^{\dagger} c_{i,B} \rangle)\\
Q_{on} &\rightarrow Q_{on} - V(\langle n_{i,A} \rangle-\langle n_{i,B} \rangle).
\end{aligned}
\end{equation}

In Fig.~\ref{exciton_correlation}, we show the $\langle n_{i,c}(1-n_{j,v}) \rangle$ correlation function for $\Omega=0.7151$ (upper panel) and $\Omega=0.7806$ (lower panel). These frequencies correspond to the two peaks in the spectrum of the upper panel of Fig.~\ref{shift_cond_2P_effect}. The interaction strength in this calculation is $V=0.03$, for which the tdMF and correlation expansion, including two-particle correlations, yield identical results. The upper panel of Fig.~\ref{exciton_correlation}, which is calculated at the excitonic peak, demonstrates that the amplitude of the electron-hole correlation in the steady-state
rapidly decreases as the distance between the sites, $i-j$, increases. The correlation for $i-j=1$ is already ten times smaller than that for $i-j=0$. This is a manifestation of the excitonic nature of this peak, confirming a locally bound electron-hole pair.
On the other hand, as shown in the lower panel of Fig.~\ref{exciton_correlation}, the electron-hole correlations at $\Omega=0.7806$ do not show such a rapid decrease as $i-j$ is increased. For this frequency, we cannot see a locally bound electron-hole pair.
These calculations clearly show that the peak at $\Omega=0.7151$ has an excitonic nature, and the peak at $\Omega=0.7806$ does not. For a larger interaction, e.g., $V=0.15$, the excitonic nature is enhanced, and the excitonic peak plays a more important role, as shown in the lower panel of Fig.~\ref{shift_cond_2P_effect}.

\begin{figure}[t]
\begin{center}
\includegraphics[width=0.73\linewidth]{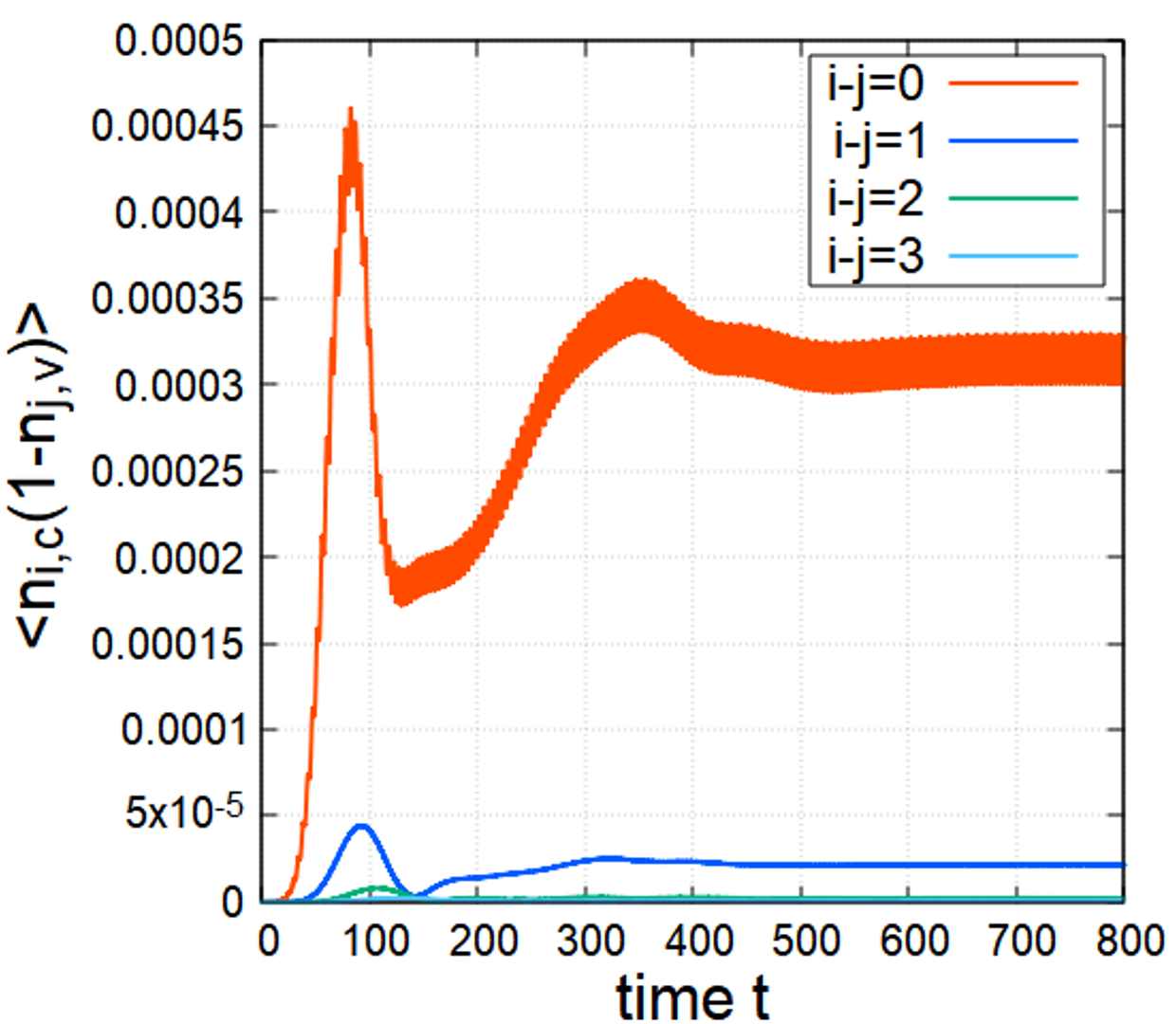}
\\
\includegraphics[width=0.73\linewidth]{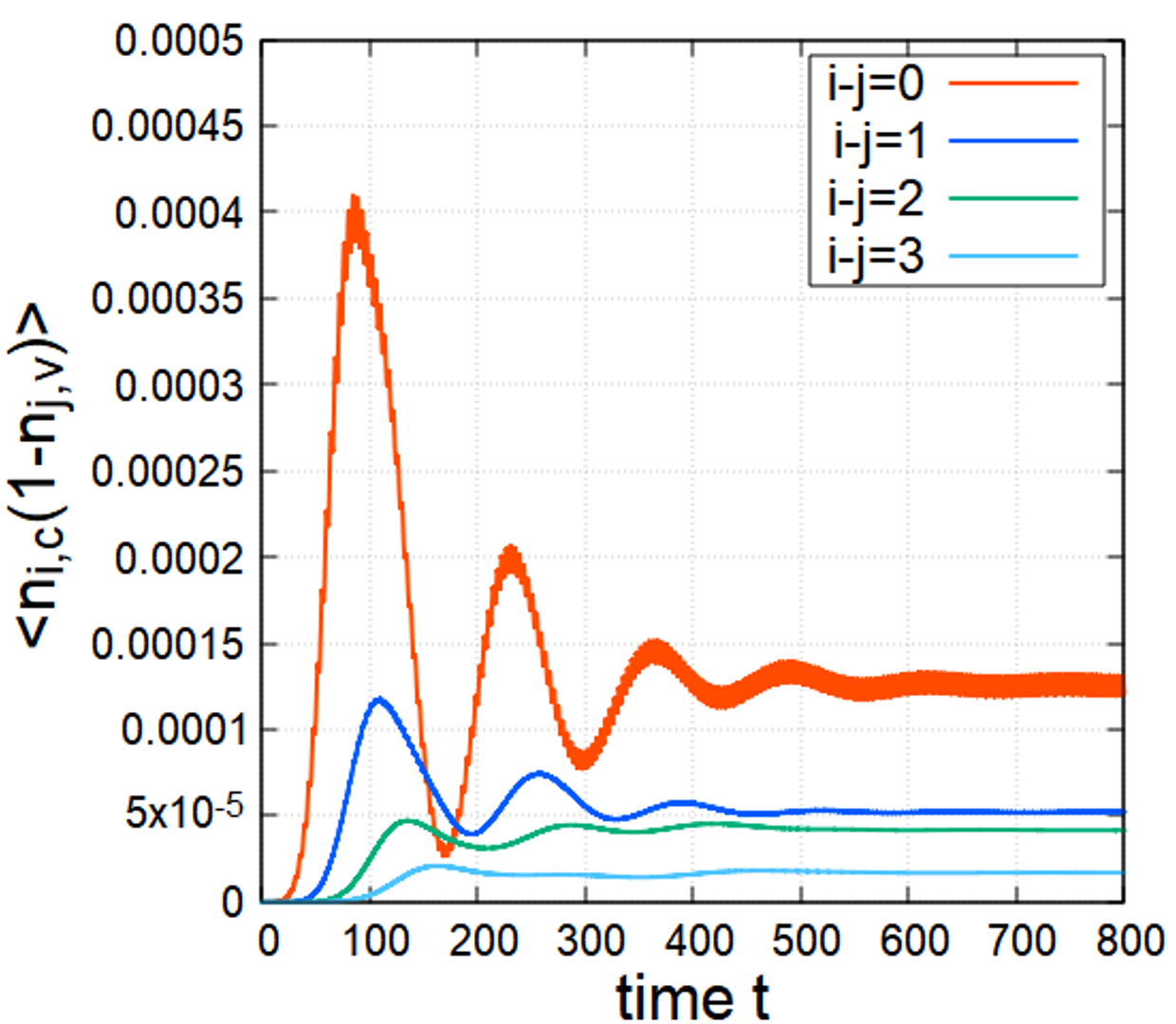}
\end{center}
\caption{Comparison of the conduction-electron-valence-hole correlations calculated in real space for different distances between electron and hole. The upper panel shows the time-resolved correlations at the excitonic peak, demonstrating a substantial enhancement of the local correlations. The lower panel shows the time-resolved correlations away from the excitonic peak, where these correlations are still comparably strong for large distances. }
\label{exciton_correlation}
\end{figure}
\bibliography{bunken}

\end{document}